\pdfoutput=1
\documentclass[12pt]{article}

\usepackage{graphicx,psfrag,epsf,color}
\usepackage{amsmath,amssymb,amsfonts}
\usepackage{array}
\usepackage{cite}
\usepackage{multirow}
\usepackage{rotating}
\usepackage{slashed}
\usepackage{bm}

\newlength{\leftstackrelawd}
\newlength{\leftstackrelbwd}
\def\leftstackrel#1#2{\settowidth{\leftstackrelawd}%
{${{}^{#1}}$}\settowidth{\leftstackrelbwd}{$#2$}%
\addtolength{\leftstackrelawd}{-\leftstackrelbwd}%
\leavevmode\ifthenelse{\lengthtest{\leftstackrelawd>0pt}}%
{\kern-.5\leftstackrelawd}{}\mathrel{\mathop{#2}\limits^{#1}}}

\newcommand{\be}{\begin{equation}}
\newcommand{\ee}{\end{equation}}
\newcommand{\bea}{\begin{eqnarray}}
\newcommand{\eea}{\end{eqnarray}}
\newcommand{\bi}{\begin{itemize}}
\newcommand{\ei}{\end{itemize}}
\newcommand{\ben}{\begin{enumerate}}
\newcommand{\een}{\end{enumerate}}
\newcommand{\bt}{\begin{tabular}}
\newcommand{\et}{\end{tabular}}

\newcommand{\mchi}{m_\chi}

\newcommand{\mt}{m_t}
\newcommand{\mh}{m_H}
\newcommand{\mW}{m_W}
\newcommand{\mZ}{m_Z}

\numberwithin{equation}{section}

\begin{document}
\allowdisplaybreaks

\begin{titlepage}

\begin{flushright}
{\small
TUM-HEP-1361/21\\
August 16, 2021    
}
\end{flushright}

\vskip1cm
\begin{center}
    {\Large \bf NLO electroweak potentials for \\[0.1cm] minimal dark matter and beyond}\\[0.2cm]
\end{center}

  \vspace{0.5cm}
\begin{center}
{\sc Kai~Urban
\\[6mm]
{\it Physik Department T31,\\
James-Franck-Stra\ss e~1, 
Technische Universit\"at M\"unchen,\\
D--85748 Garching, Germany}}
\\[0.3cm]
\end{center}

\vspace{0.6cm}
\begin{abstract}
\vskip0.2cm\noindent
We calculate the one-loop correction to the static potential induced by $\gamma$, $W$ and $Z$-exchange at tree-level for arbitrary heavy standard model multiplets. We find that the result obeys a ``Casimir-like" scaling, making the NLO correction to the potential a ``low-energy" property of the electroweak gauge bosons. Furthermore, we discuss the phenomenology of the NLO potentials, the analytically known asymptotic limits and provide fitting functions in position space for easy use of the results.
\end{abstract}
\end{titlepage}

\section{Introduction}
\label{sec:introduction}
In the context of (electro-)weakly interacting dark matter (DM), the static potential plays a crucial role in substantially modifying the annihilation cross-section through the Sommerfeld effect in indirect detection and relic abundance calculations \cite{Hisano:2004ds,Hisano:2006nn}. Due to the finite range nature of the Yukawa potentials arising from $W,Z$-exchange, resonant features associated with zero-energy bound states that enhance the cross-sections by up to several orders of magnitude are observed in the mass spectrum. The size of the enhancement effect has made the accurate calculation of the leading-order (LO) Sommerfeld effect an essential ingredient in many WIMP DM calculations in the literature, both for minimal DM models \cite{Cirelli:2007xd} and the MSSM \cite{Beneke:2014gja,Beneke:2014hja}.

In recent years, the calculation of the annihilation cross-section in DM indirect detection $\chi \chi \to \gamma + X$ has progressed to include electroweak (EW) Sudakov logarithms and their resummation to all orders \cite{Baumgart:2014vma,Bauer:2014ula,Ovanesyan:2014fwa,Ovanesyan:2016vkk,Baumgart:2017nsr,Baumgart:2018yed,Beneke:2018ssm,Beneke:2019vhz,Beneke:2019gtg}. The precision of these calculations has reached the percent level for the wino and Higgsino model, naturally facilitating the question of the size of next-to-leading order (NLO) electroweak corrections to the non-relativistic Sommerfeld effect expected to be the dominant source of theoretical uncertainty. Possible corrections to the Sommerfeld effect are the mass-splittings between the particles in the multiplet after electroweak symmetry breaking (EWSB) to two-loops, which are known for the simplest multiplets \cite{Yamada:2009ve,Ibe:2012sx,McKay:2017xlc}, and the NLO EW potentials. For the case of wino DM, the NLO potentials were first calculated in \cite{Beneke:2019qaa,Beneke:2020vff} and shown to yield sizeable corrections, of similar size than the Sudakov resummation, to the indirect detection prediction due to shifts of the zero-energy bound state energies \cite{Beneke:2019qaa}. Also, for relic density calculations, corrections are non-negligible though not quite as significant due to a complex interplay between high and low temperature regimes \cite{Beneke:2020vff}.  

Going away from the simple test case presented by wino DM, the question arises on how these results can be generalized to arbitrary multiplets, possibly allowing for non-zero hypercharge. For wino DM, channels with the same tree-level potential (up to possible signs) were observed to receive the same NLO correction \cite{Beneke:2020vff}. In addition, the fact that the standard model (SM) self-energies tied to the tree-level exchange bosons have a highly complex gauge parameter dependence that needs to be cancelled by other topologies suggests a universality to be uncovered. 

For massless gauge theories, it is known for a long time, that the correction to the static potential obeys a Casimir scaling \cite{Ambjorn:1984mb}. In the case of QCD, the Casimir scaling of the static potential is only violated at the 3-loop order \cite{Anzai:2010td}. At the one-loop order the universality of the potential correction is easiest seen by choosing a Coulomb gauge formulation in which only self-energies contribute to the NLO correction \cite{Appelquist1978}. In spontaneously broken gauge theories, it is not intuitively clear that the conceptually interesting features of Coulomb gauge are also applicable. Already in the 1980's, a generalized Coulomb gauge for a spontaneously broken gauge theory with massless and massive gauge bosons was formulated (though not for the SM), with similar features than in the massless case \cite{Kreuzer:1986nu}. The existence of a Coulomb gauge formulation thereby suggests that the correction in the SM also has to obey a ``Casimir-like" scaling. 

In this paper, we calculate the NLO correction to the static potential between two SM multiplet particles, not necessarily part of the same multiplet. The NLO correction is only tied to the particle content and gauge group of the SM. In this sense, it represents a ``low-energy" property of the SM gauge bosons, i.e., for exchange momenta of the order of the EW scale $|\mathbf{k}| \sim \mW$. In addition, EWSB and the interplay of massless and massive gauge boson exchange allow for unique infrared behaviour for special linear combinations of couplings that have no analogue in the QCD literature.

The paper is structured as follows: First, we provide a short overview of the effective field theory setup and discuss the construction of the general tree-level potentials for an arbitrary SM multiplet of TeV-scale mass. Afterwards, we provide the NLO potentials in all possible channels. We omit many of the technical details, e.g., the calculation of the one-loop topologies, which are provided in \cite{Beneke:2020vff,Urban:2021evz} and focus on the phenomenology of the NLO potentials.  Subsequently, we present a detailed analysis of all channels that have not appeared in the context of the wino NLO potentials in \cite{Beneke:2019qaa,Beneke:2020vff,Urban:2021evz}, provide accurate fitting functions for easy use, and some general comments before we conclude. 

\section{EFT setup and tree-level potentials}
\label{sec:EFTsetup}
We follow the effective theory setup as outlined for non-relativistic heavy WIMPs in \cite{Beneke:2012tg,Hellmann:2013jxa,Beneke:2014hja} that was constructed in analogy to the respective EFTs for QED and QCD \cite{Pineda:1997bj,Beneke:1998jj,Beneke:1999qg,Brambilla:1999xf,Bodwin:1994jh}. We start from the non-relativistic effective theory
\begin{align}
    \mathcal{L}_{\rm NRDM} &= \chi^\dagger(x) \left(i D^0 +
    \frac{\mathbf{D}^2}{2 \mchi}\right) \chi(x)\,,
    \label{eq:LNRDM}
\end{align}
where $\chi$ is an arbitrary spin multiplet field charged under SU(2)$_L$ and/or hypercharge. To stay close to the previous literature \cite{Beneke:2012tg,Beneke:2014gja,Beneke:2019qaa,Beneke:2020vff} on non-relativistic effective theories for electroweak DM at the TeV scale, we use in the following the term DM synonymous to non-relativistic particles of mass $\mchi \gg \mZ$, if not explicitly stated otherwise. In fact, calculating the potentials to NLO, the assumption of DM is not needed in any step.

Technically, the potentials are a matching coefficient onto a non-local four-fermion operator from the non-relativistic theory to the potential non-relativistic theory. The necessary Lagrangian pieces to obtain NLO non-relativistic accuracy are given by
\begin{align}
\label{eq:LPNRDM}
\mathcal{L}_{\rm PNRDM} &= \sum_{i} \chi_{vi}^\dagger(x) \left(i
D^0_i(t,\mathbf{0}) - \delta m_i+ \frac{\bm{\partial}^2}{2\mchi} \right)\chi_{vi}(x) \nonumber\\ 
                        &-\,\sum_{\{i,j\},\{k,l\}} \int d^3\mathbf{r}\,
V_{(ij),(kl)}(r)\, \chi_{vk}^\dagger(t,\mathbf{x})
\chi_{vl}^\dagger(t,\mathbf{x}+\mathbf{r}) \chi_{vi}(t,\mathbf{x})
\chi_{vj}(t,\mathbf{x}+\mathbf{r}) \, ,
\end{align}
where $\delta m_i$ is the mass splitting between members of the multiplet due to radiative corrections and, in typical cases, is of the order of the kinetic energy. The NLO correction to the potentials $V_{(ij)(kl)}$ arises from the soft region in the method of regions expansion~\cite{Beneke:1997zp}. For a detailed discussion of the power counting, the possible further terms in the Lagrangians, and other aspects, see \cite{Beneke:2020vff}.

\subsection{Tree-level potentials for arbitrary SM representations}
The above allows us to extract the tree-level potentials for arbitrary SM representations by matching the tree-level exchange in the potential region \cite{Beneke:1997zp}. The necessary electroweak Feynman rules for the emission from a static heavy particle depend only on the representation under ${\rm SU(2)}_L$ and ${\rm U(1)}_Y$. We select to work in the charge basis where $T^3_R$ is diagonal, i.e., the electric charge is given by $Q=T^3_R + Y$. Additionally, there is some freedom in choosing the entries of $T^\pm_R$, which are only fixed up to an arbitrary phase for each multiplet component. We choose the entries of $T^\pm_R$ real, which is also consistent with the minimal DM requirement of having the same SU(2) invariant mass term for all components of the multiplet \cite{Cirelli:2007xd}. The generic spin-$j$ representation for this convention is then constructed using \cite{Georgi:1999wka}
\begin{align}
    \left\langle j, m_1 \middle| T^3_R \middle| j , m_2 \right\rangle &= m_2\, \delta_{m_1,m_2} \nonumber \\
    \left\langle j, m_1 \middle| T^+_R \middle| j , m_2 \right\rangle &= \sqrt{\frac{(j+m_2+1)(j-m_2)}{2}} \, \delta_{m_1,m_2 +1} \nonumber \\
    \left\langle j, m_1 \middle| T^-_R \middle| j , m_2 \right\rangle &= \sqrt{\frac{(j+m_2)(j-m_2+1)}{2}} \, \delta_{m_1,m_2-1} \, .
\end{align}
Using these generators, the tree-level potential for minimal DM \cite{Cirelli:2005uq,Cirelli:2007xd} and even more generally arbitrary SM representations (not necessarily including a DM candidate) can be constructed. The only assumption to be made is that the heavy particles that act as static sources $\mchi \gg \mZ$ are charged under the EW sector of the SM.

For the (off-diagonal) tree-level potential due to the exchange of a $W$-boson, we find
\begin{align}
  V^{W}_{(ij),(kl)} &=(-1)^{n_Q} \frac{4\pi \alpha_2}{\mathbf{k}^2 + m_W^2}\left( T^+_{R,ik} T^-_{R,jl} + T^-_{R,ik} T^+_{R,jl}\right) \nonumber \\ 
                    &=(-1)^{n_Q} V^W_{\rm tree} \left( T^+_{R,ik} T^-_{R,jl} + T^-_{R,ik} T^+_{R,jl}\right) \, ,
\end{align}
where $n_Q = 0$ if the $|Q_i|+|Q_j| = |Q_k| + |Q_l|$ and $n_Q =1$ otherwise. $(-1)^{n_Q}$ expresses the possible minus sign in the Feynman rule in the employed particle/particle convention for the non-relativistic Lagrangian \cite{Beneke:2012tg}. In the particle/anti-particle convention oftentimes used in NRQCD, the sign arises due to ``fermion flow''. Let us note that the factor is mainly used to allow compact results, as many expressions differ simply by a sign. One can also avoid using this factor at all, however, at the cost of introducing two $W$-exchange potentials to cover all possible cases. Given the NLO potentials discussed later, we find using a single factor to express this possible sign more convenient. In addition, let us remark that for the group factor $T^+_R T^-_R + T^-_R T^+_R$, only one term is non-zero for a fixed channel.

Similarly, we construct the photon and $Z$-exchange potentials
\begin{align}
    V^{\gamma/Z}_{(ij),(ij)} &= + \frac{4 \pi \alpha}{\mathbf{k}^2} (T^3_{R,ii} + Y_i) (T^3_{R,jj} + Y_j) + \frac{4 \pi \alpha}{\mathbf{k}^2 + m_Z^2}  \frac{c_W^2 T^3_{R,ii} - s_W^2 Y_i}{s_W c_W}\cdot \frac{c_W^2 T^3_{R,jj} - s_W^2 Y_j}{s_W c_W}  \nonumber \\
                             &= \left(\frac{4 \pi \alpha}{\mathbf{k}^2} + \frac{4 \pi \alpha}{\mathbf{k}^2 + m_Z^2} \frac{c_W^2}{s_W^2}\right) T^3_{R,ii} T^3_{R,jj} + \left(\frac{4 \pi \alpha}{\mathbf{k}^2} -\frac{4 \pi \alpha}{\mathbf{k}^2 + m_Z^2} \right) (T^3_{R,ii} Y_j+T^3_{R,jj} Y_i)  \nonumber \\
                             &\quad +\left(\frac{4 \pi \alpha}{\mathbf{k}^2} + \frac{4 \pi \alpha}{\mathbf{k}^2 + m_Z^2} \frac{s_W^2}{c_W^2}\right) Y_i Y_j \nonumber \\[0.1cm]
                             &= V^{T3T3}_{\rm tree} \, T^3_{R,ii} T^3_{R,jj} + V^{T3Y}_{\rm tree}(T^3_{R,ii} Y_j+T^3_{R,jj} Y_i) + V^{YY}_{\rm tree} \, Y_i Y_j \nonumber \\[0.1cm]
                             &= V^{T3T3}_{(ij)(ij)} + V^{T3Y}_{(ij)(ij)} + V^{YY}_{(ij)(ij)} \, ,
\end{align}
where we fixed $i=k,j=l$, as hypercharge $Y$ and $T^3_R$ are diagonal matrices. The hypercharges for particles $i,j$ are distinguished to cover antiparticles for which $Y_i = -Y_j$. Although, we assume the same SU(2) representation $R$ for both sources, to not further clutter notation, all results, including NLO potentials also hold if the sources are part of different multiplets, as we explicitly checked.

We identify three gauge-invariant combinations to which we will consider the NLO correction separately. The combinations $T^3 T^3$ and $YY$ are gauge-invariant, as one can write down models with vanishing SU(2) charge and non-zero hypercharge and vice versa. As these two combinations need to be separately gauge-invariant, also the linear combination $T^3Y$ is gauge-invariant, even though it only appears if both $T^3T^3$ and $YY$ are non-zero.\footnote{As long as the potential is between particles of the same multiplet, $T^3 Y$ only appears if both other combinations are non-zero. In the case of different multiplets, situations may occur, where both $T^3_R T^3_{R^\prime}$ and $Y_R Y_{R^\prime}$ are zero, whilst $T^3_R Y_{R^\prime}$ is not, e.g., if one static source has zero hypercharge and the other non-zero hypercharge but vanishing SU(2) charge. However, the following analysis is not affected by this consideration.}

\section{NLO electroweak potentials}
\label{sec:NLOpotentials}
To calculate the NLO potential correction, we consider all possible one-loop topologies with general SU(2) and hypercharge factors. Using general group identities for arbitrary SU(2) representations (see, e.g., \cite{vanRitbergen:1998pn}), all topologies are reduced till only two SU(2) or hypercharge factors are present. Together with the necessary loop expressions, which are discussed in \cite{Beneke:2020vff} (Appendix A) in Feynman and general covariant $R_\xi$-gauge, and the on-shell renormalization scheme (Section 3.1.1) of \cite{Beneke:2020vff} the NLO correction can be assembled. For brevity, we point to \cite{Beneke:2020vff} for the notation, topologies and other subtleties in connection with loop integrals and on-shell renormalization scheme.

\subsection{Off-diagonal $W$-boson exchange}
\begin{figure}[t!] \centering
    \includegraphics[width=\textwidth]{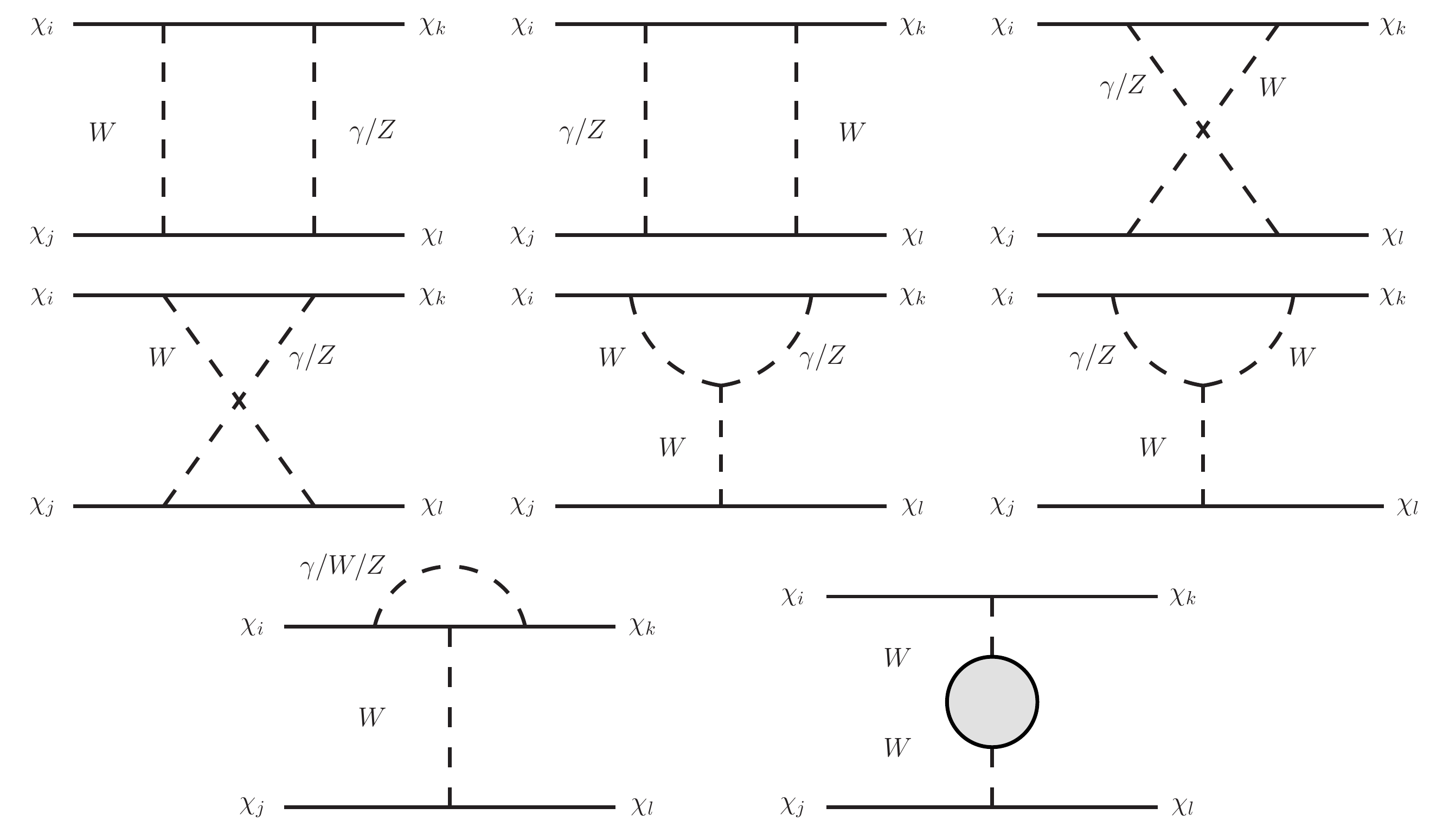}
    \caption{Relevant one-loop topologies correcting the tree-level $W$-exchange channel $\chi_i \chi_j \to \chi_k \chi_l$ channel excluding wave-function and counterterm topologies. Also symmetric diagrams, e.g., vertex corrections at the lower line are not explicitly shown.} \label{fig:Wexchange_topologies} 
  \end{figure}

Let us begin by examining the correction corresponding to the tree-level off-diagonal $W$-boson exchange contribution. After assembling all topologies and counterterms, as schematically shown in Figure~\ref{fig:Wexchange_topologies}, we are left with the correction
\begin{align}
    &\delta V^{W}_{(ij),(kl)} = V^W_{(ij),(kl)} \left[2 \delta Z_e - 2 \frac{\delta s_W}{s_W} + \frac{\Sigma_T^{WW}(-\mathbf{k}^2) - \delta m_W^2}{\mathbf{k}^2 + m_W^2}+2 \left(I^{W\gamma}_{\rm 3\,gauge} + I^{WZ}_{\rm 3\, gauge}\right)\right] \nonumber \\
                             &\quad +(-1)^{n_Q+1} (T^+_{R,ik} T^-_{R,jl} +T^-_{R,ik} T^+_{R,jl}) \left[ I_{\rm box}\left(\alpha_2  ,m_W;\alpha ,0\right) + I_{\rm box}\left(\alpha_2 ,m_W;\alpha \frac{c_W^2}{s_W^2}, m_Z\right) \right] \nonumber \\
                             &\quad + \left. \delta V^W_{(ij),(kl)} \right|_{\rm WF/vertex} \, .
\end{align}
The first line gives the contribution trivially proportional to the tree-level potential, as it corresponds to the electric charge and Weinberg angle counterterms, self-energy, $W$-boson mass counterterm insertion, and the triple vertex topologies. In the second line the box topologies contribute. As only one $W$-boson can be part of these corrections to have a net charge change of one between the two static sources, it is also easy to verify, that this contribution is proportional to the tree-level group factor. Finally, in the last line there is symbolically the wave-function and vertex correction. Both wave-function, as well as vertex correction are proportional to further group factors. In the linear combination of the two, however, only a contribution proportional to the tree-level group factor remains. The resulting expression for the one-loop correction is gauge-invariant and finite, as we explicitly checked. Furthermore, in the limits $\mW \to \mZ$ and hence $s_W \to 0, c_W \to 1$, we recover previously known expressions for the singlet potential of a Higgsed SU(2) theory \cite{Laine:1999rv,Schroder:1999sg} (similarly for $T^3 T^3$ below). More precisely, we compared the unrenormalized potential, as the renormalization was not fully specified in \cite{Laine:1999rv}.

Therefore, the one-loop correction to the $W$-boson exchange potentials can be written as
\begin{align}
  V^{W,{\rm NLO}}_{(ij),(kl)} = (-1)^{n_Q} \left( T^+_{R,ik} T^-_{R,jl} + T^-_{R,ik} T^+_{R,jl}\right) \left(V^W_{\rm tree} + \delta V^W_{\rm 1-loop}\right) \, ,
  \label{eq:CasimirWexchange}
\end{align}
meaning the potential correction is universal. Like in QED and QCD, where the potential correction scales with the Casimir operator of the representation, we find a ``Casimir-like'' scaling analogously also in the spontaneously broken EW theory. In this sense, the correction is a ``low-energy'' property of the SM $W$-bosons.

\subsection{Diagonal photon and $Z$-boson exchange}
In an analogous fashion, we can assemble the correction for the diagonal channels $(ij) \to (ij)$, i.e., corresponding to tree-level photon and/or $Z$-boson exchange. Again we find that we can reduce all contributions to the tree-level factors. As mentioned for the tree-level potentials above, we split the corrections into separately gauge-invariant contributions. The correction proportional to $T^3 T^3$ is given by
\begin{align}
    \delta V^{T3T3}_{(ij),(ij)} &= + \frac{4 \pi \alpha}{\mathbf{k}^2} \left(T^3_{R,ii} T^3_{R,jj}\right)  \left(2 \delta Z_e + \frac{\Sigma^{\gamma \gamma}_T(-\mathbf{k}^2)}{\mathbf{k}^2}\right)\nonumber \\
                                                                        &\quad+ \frac{4 \pi \alpha}{\mathbf{k}^2 + m_Z^2}\left(T^3_{R,ii} T^3_{R,jj}\right) \frac{c_W^2}{s_W^2}\left(2 \delta Z_e +2 \frac{\delta c_W}{c_W} -2 \frac{\delta s_W}{s_W} + \frac{\Sigma^{ZZ}_T(-\mathbf{k}^2)-\delta m_Z^2}{\mathbf{k}^2+m_Z^2}\right) \nonumber \\
                                                                        &\quad + \frac{4 \pi \alpha}{\mathbf{k}^2 (\mathbf{k}^2 + m_Z^2)}\left(T^3_{R,ii} T^3_{R,jj}\right) \frac{-2 c_W}{s_W} \Sigma^{\gamma Z}_T(-\mathbf{k}^2) + V^{T3T3}_{(ij),(ij)} \left(2 I^{WW}_{\rm 3 \, gauge}\right)  \nonumber \\
                                                                        &\quad + \left. \delta V^{T3T3}_{(ij),(ij)} \right|_{\rm vertex/WF}-T^3_{R,ii} T^3_{R,jj} I_{\rm box}\left(\alpha_2,m_W;\alpha_2,m_W\right) \, .
                                                                        \label{eq:CasimirgamZ}
\end{align}
Similar to the off-diagonal correction discussed above, the first three lines give the corrections trivially proportional to the tree-level group factor from counterterms, self-energies and the triple gauge vertex contribution. In the last line, the wave-function renormalization and vertex correction cancel such that only a universal term proportional to the tree-level group factor is left over. Finally, the box and crossed-box diagrams involving $W$-bosons are also proportional to $T^3_R T^3_R$, which is easily verified by writing out the group factors, accounting for the relative minus sign of the crossed box to the box diagram, and in addition using $\left[T^+_R , T^-_R \right]_{ii} = - T^3_{R,ii}$.

For the tree-level $T^3 Y$ linear combination, the correction reads
\begin{align}
 \delta V^{T3Y}_{(ij),(ij)} &= + \frac{4 \pi \alpha}{\mathbf{k}^2} \left(T^3_{R,ii} Y_j + T^3_{R,jj} Y_i\right)  \left(2 \delta Z_e + \frac{\Sigma^{\gamma \gamma}_T(-\mathbf{k}^2)}{\mathbf{k}^2}\right)\nonumber \\
                                                                        &\quad- \frac{4 \pi \alpha}{\mathbf{k}^2 + m_Z^2} \left(T^3_{R,ii} Y_j + T^3_{R,jj} Y_i\right) \left(2 \delta Z_e + \frac{\Sigma^{ZZ}_T(-\mathbf{k}^2)-\delta m_Z^2}{\mathbf{k}^2+m_Z^2}\right) \nonumber \\
                                                                        &\quad + \frac{4 \pi \alpha}{\mathbf{k}^2 (\mathbf{k}^2 + m_Z^2)} \left(T^3_{R,ii} Y_j + T^3_{R,jj} Y_i\right)\left(\frac{s_W}{c_W}-\frac{c_W}{s_W}\right) \Sigma^{\gamma Z}_T(-\mathbf{k}^2) \nonumber \\
                                                                        &\quad +\left. \delta V^{T3Y}_{(ij),(ij)} \right|_{\rm vertex/WF} + V^{T3Y}_{(ij),(ij)} I^{WW}_{\rm 3 \, gauge} \, .
\end{align}
Note that in this case, and also in the $YY$ case below, the box contributions cancel against their crossed box counterparts. In the diagonal channel, there is always the possibility of pure photonic and pure $Z$ box contributions that cancel naturally. Furthermore, the mixed photon/$Z$ boxes cancel, as the photon and $Z$-boson commute.  Therefore, only $W$-boson boxes contribute in diagonal channels, which are pure SU(2) contributions and therefore have the $T^3_R T^3_R$ structure discussed above. Furthermore, in this case, and for $YY$ below, the wave-function and vertex contribution is proportional to the tree-level potential factor. All other topologies are trivially proportional to the tree-level potential. For $YY$, the result reads
\begin{align}
\delta V^{YY}_{(ij),(ij)} &= + \frac{4 \pi \alpha}{\mathbf{k}^2} \left(Y_i Y_j\right)  \left(2 \delta Z_e + \frac{\Sigma^{\gamma \gamma}_T(-\mathbf{k}^2)}{\mathbf{k}^2}\right)\nonumber \\
                                                                        &\quad+ \frac{4 \pi \alpha}{\mathbf{k}^2 + m_Z^2} \frac{s_W^2}{c_W^2}\left(Y_i Y_j\right) \left(2 \delta Z_e -2 \frac{\delta c_W}{c_W} +2 \frac{\delta s_W}{s_W} + \frac{\Sigma^{ZZ}_T(-\mathbf{k}^2)-\delta m_Z^2}{\mathbf{k}^2+m_Z^2}\right) \nonumber \\
                                                                        &\quad + \frac{4 \pi \alpha}{\mathbf{k}^2 (\mathbf{k}^2 + m_Z^2)}\left(Y_i Y_j\right) \frac{2 s_W}{c_W} \Sigma^{\gamma Z}_T(-\mathbf{k}^2) + \left. \delta V^{YY}_{(ij),(ij)}\right|_{\rm vertex/WF} \, .
\end{align}
For the pure hypercharge contribution, also the triple vertex contribution is non-existent. The origin of the triple gauge interaction is the $W^a W^b W^c$ interaction in the unbroken theory.  Only one of these bosons can be $W^3$ that in the broken theory becomes photon or $Z$-boson. Therefore, only once a mixing of SU(2) and U(1) is possible, meaning only $T^3 T^3$ and $T^3 Y$ have a triple gauge contribution. We checked that the poles cancel for each of the three gauge-invariant linear combinations and that the result is independent of the gauge-fixing parameters $\xi_\gamma, \xi_Z$, and $\xi_W$.

As for the off-diagonal contribution above, in the diagonal channel with tree-level photon and/or $Z$-exchange, we find
\begin{align}
  V^{\gamma/Z, {\rm NLO}}_{(ij),(ij)} &= V^{T3T3}_{(ij),(ij)} + \delta V^{T3T3}_{(ij),(ij)} + V^{T3Y}_{(ij),(ij)}+\delta V^{T3Y}_{(ij),(ij)}+V^{YY}_{(ij),(ij)}+\delta V^{YY}_{(ij),(ij)} \nonumber \\
                                       &= T^3_{R,ii} T^3_{R,jj}  \left(V^{T3T3}_{\rm tree} +\delta V^{T3T3}_{\rm 1-loop} \right) +(T^3_{R,ii} Y_j+T^3_{R,jj} Y_i) \left(V^{T3Y}_{\rm tree}+\delta V^{T3Y}_{\rm 1-loop}\right) \nonumber \\
   &\quad + Y_i Y_j\left(V^{YY}_{\rm tree}+\delta V^{YY}_{\rm 1-loop}\right) \, ,
\end{align}
meaning that also in this channel the one-loop correction is universal and only dependent on the tree-level gauge boson exchange.

\subsection{One-loop correction for vanishing tree-level potential}
There are, in principle, two further possibilities for possible channels, induced at one-loop as their tree-level potential vanishes. Channels with charge change two between the static sources, e.g., $(00) \to ((++)\, (--))$ in the SU(2) quintuplet, or channels with net charge change zero between the sources, but vanishing tree-level potential, e.g., $(00) \to (00)$ for the pure wino.

In both cases, the correction at one-loop vanishes. Intuitively, this is also expected, as the only topologies that can induce such a potential are box and crossed box diagrams. Given their very intricate gauge parameter dependence, which can only be cancelled by a delicate linear combination of self-energies, counterterms and so on, the box and crossed box diagrams were expected to cancel amongst themselves to maintain gauge invariance. We also checked the cancellation explicitly.

Another way to think of this cancellation is to note that Coulomb gauge can also be formulated for spontaneously broken non-Abelian gauge theories \cite{Kreuzer:1986nu}. In Coulomb gauge, all soft emissions connected to the heavy static source vanish. Therefore, only the potential region of the box diagram, which is an iteration of the tree-level potential and the Coulomb gauge SM self-energies and counterterms, may contribute. However, the self-energies and counterterm insertions are necessarily proportional to the tree-level potential, which in the cases at hand vanishes, meaning that there are no induced channels at one-loop. The Coulomb gauge argument also motivates the ``Casimir-like'' scaling, as one could also phrase the result in terms of Coulomb gauge SM self-energies with all other terms vanishing.

To summarize: for hypercharge $Y=0$ independent of the SU(2) representation, two potentials, one off-diagonal due to $W$-exchange and one diagonal due to $\gamma/Z$-exchange appear. The one-loop correction to both potentials is proportional to the tree-level group factor and thereby presents the analogue of Casimir scaling in QCD. The results already appear in the literature for the wino model with corresponding group factors \cite{Beneke:2019qaa,Beneke:2020vff}. For non-zero hypercharge $Y\neq 0$, two further potentials occur that can be assembled from known results for PNRDM loop integrals (for details, see Appendix~A of \cite{Beneke:2020vff}). The correction to these additional potentials in the diagonal channel is also proportional to the tree-level group factor, meaning they also obey Casimir-like scaling. Furthermore, underlining the scaling behaviour of the NLO correction, we find that there are no one-loop induced potentials in channels with a vanishing tree-level contribution.

\section{Analysis of the different channels}
\label{sec:channels}
In this section, we discuss the behaviour of the NLO potential in the asymptotic regions, where analytic results in position space can be extracted and thereby, further strong checks on the results are possible. We omit the discussion of the correction to off-diagonal $W$-boson exchange and the $T^3 T^3$ linear combination for $\gamma/Z$-exchange, as these appear already for wino DM \cite{Beneke:2019qaa,Beneke:2020vff} and are extensively discussed in Section~3.3 of \cite{Beneke:2020vff}. Therefore, the focus is on $T^3 Y$ and $YY$, and particular linear combinations of $T^3$ and $Y$ with special long-distance behaviour. To make the origin of the leading behaviours and the various factors more transparent, we discuss the asymptotic corrections for three separately gauge-invariant contributions -- light fermions, third-generation quarks, and electroweak sector.

Furthermore, we provide fitting functions that approximate the potential correction to permille level accuracy, making them suited, e.g., for Sommerfeld enhancement or bound state calculations in minimal DM models. Finally, we comment on the various dependencies on the renormalization scheme, top quark mass, and further subtleties.

All numerical values, unless stated otherwise are obtained with the input parameters: $\alpha = \alpha_{\rm os}(\mZ) = 1/128.943$ for the electromagnetic coupling at the $Z$-mass, the gauge-boson masses $\mW = 80.385 \,{\rm GeV}$ and $\mZ = 91.1876 \,{\rm GeV}$, that via on-shell relations determine the weak couplings and mixing angles $c_W = \mW / \mZ$. Furthermore, the value for Higgs boson $m_H = 125 \,{\rm GeV}$ and top quark mass $m_t = 173.1 \,{\rm GeV}$ are used. All other fermions of the SM are taken massless. For practical purposes the uncertainty on all parameters, except the top mass dependence discussed below, is negligible.

\subsection{Asymptotic behaviour}\label{sec:asymain}
\subsubsection{$T^3 Y$ - linear combination}
\begin{figure}[t!] \centering
    \includegraphics[width=0.7\textwidth]{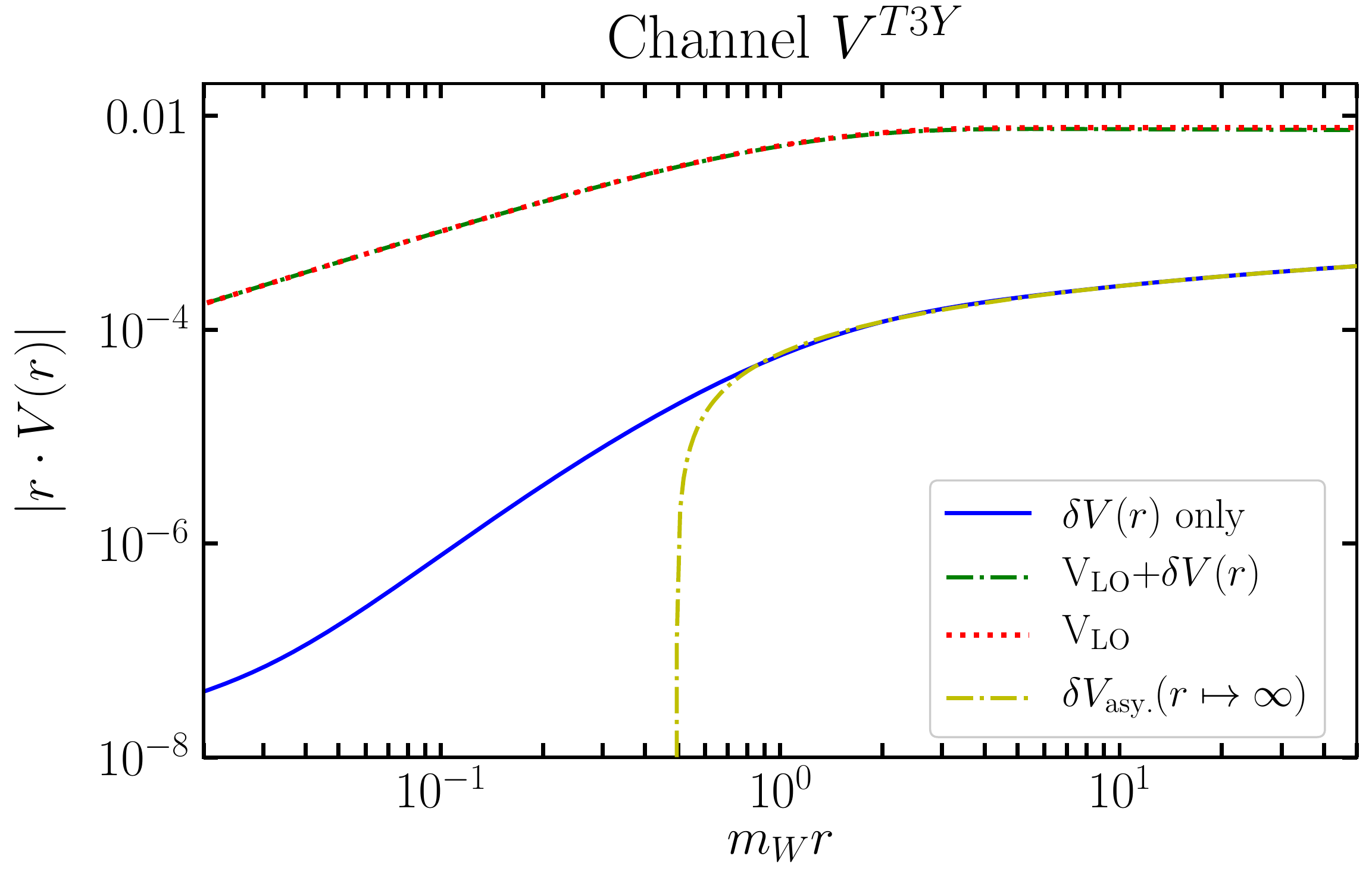}
    \hspace*{-0.51cm}\includegraphics[width=0.735\textwidth]{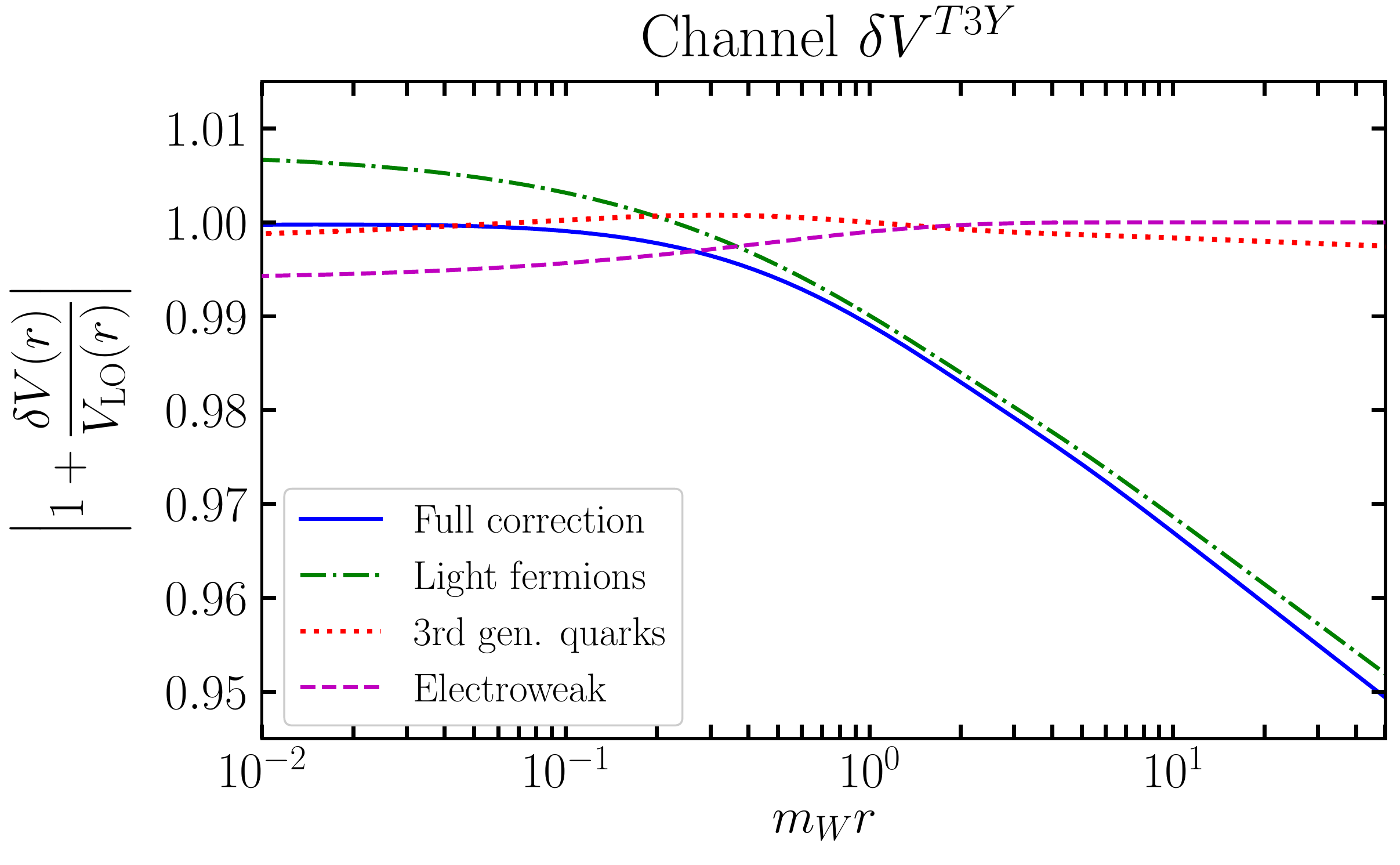}
    \caption{The NLO correction to the tree potential $V_{\rm tree}^{T3Y}$. The upper panel shows the modulus of the potential $|r \cdot V(r)|$ for the LO and NLO potential, the NLO contribution only, and the large-distance asymptotic behaviour. In the lower panel, we show the ratio of the full NLO potential to the LO potential (blue solid) and separately for the three gauge invariant pieces identified in the text (other curves).} \label{fig:PotT3Y} 
  \end{figure}
  
In Figure~\ref{fig:PotT3Y}, we show the potential correction relative to the tree-level potential for $V^{T3Y}_{\rm tree}$. For small distances, the Coulombic behaviour of photon and $Z$ tree-level potential cancels, as SU(2) and U(1) disentangle in the UV. Similarly, for the NLO correction, we find
\begin{align}
  \delta V^{T3 Y} (r \to 0) = {\rm const.}
  \label{eq:T3Yto0}
\end{align}
which also holds for each gauge-invariant sub piece individually, as visible in the lower panel of Figure~\ref{fig:PotT3Y} by the constant ratio between LO and NLO potential. The origin of the constant term becomes clear when examining the tree-level linear combination in the method of regions expansion \cite{Beneke:1997zp}. The Fourier transformation from momentum to position space of a Yukawa potential has two regions for $r \to 0$. The first region is given by $\frac{1}{r}\sim |\mathbf{k}| \gg \mZ$, which in dimensional regularization can be evaluated order by order to
\begin{align}
  \tilde{\mu}^{2\epsilon}  \int \frac{d^{d-1} \mathbf{k}}{(2 \pi)^{d-1}}\, e^{i \mathbf{k} \cdot \mathbf{r}} \frac{4 \pi \alpha}{\mathbf{k}^2 + \mZ^2} &= \tilde{\mu}^{2 \epsilon}\int \frac{d^{d-1} \mathbf{k} }{(2 \pi)^{d-1}}\, e^{i \mathbf{k}\cdot \mathbf{r}} \,  \sum_{n=0}^{\infty}\frac{4 \pi \alpha}{\mathbf{k}^2} \left(- \frac{\mZ^2}{\mathbf{k}^2}\right)^n \nonumber \\[0.1cm]
                                                                                                                                                                 &\leftstackrel{\frac{1}{r} \sim |\mathbf{k}|\gg \mZ}{\approx}\sum_{n=0}^{\infty}\, \tilde{\mu}^{2 \epsilon}\int \frac{d^{d-1} \mathbf{k} }{(2 \pi)^{d-1}}\, e^{i \mathbf{k}\cdot \mathbf{r}} \,  \frac{4 \pi \alpha}{\mathbf{k}^2} \left(- \frac{\mZ^2}{\mathbf{k}^2}\right)^n \nonumber \\
                                                                                                                        &= \frac{\alpha}{r} \cosh(\mZ r) \ \stackrel{r \to 0}{\longrightarrow}\ \frac{\alpha}{r}\left(1 + \frac{1}{2} \mZ^2 r^2 + \mathcal{O}(\mZ^4 r^4)\right)\, ,
\end{align}
and reproduces all the attractive contributions of the Yukawa potential, particularly the leading Coulomb behaviour for $r \to 0$, that cancels against the photonic term for the $T^3 Y$ linear combination. However, to capture the screening due to the exchange boson mass, the second region that encapsulates all gauge boson mass-dependent contributions, which is given by $|\mathbf{k}|\sim \mZ \ll \frac{1}{r}$ needs to be considered. Here the gauge-boson propagator cannot be expanded. Nevertheless, we can perform an expansion in $|\mathbf{k} \cdot \mathbf{r}|\ll 1$. The resulting integrals are power-like divergent and require regularization. To evaluate the second region, we conveniently choose dimensional regularization and find
\begin{align}
  \tilde{\mu}^{2 \epsilon} \int \frac{d^{d-1} \mathbf{k} }{(2 \pi)^{d-1}}\, e^{i \mathbf{k} \cdot \mathbf{r}} \frac{4 \pi \alpha}{\mathbf{k}^2 + \mZ^2} &= \tilde{\mu}^{2 \epsilon} \int \frac{d^{d-1} \mathbf{k} }{(2 \pi)^{d-1}}\,\frac{4 \pi \alpha}{\mathbf{k}^2 + \mZ^2}\,\sum_{n=0}^{\infty} \frac{\left(i \mathbf{k} \cdot \mathbf{r}\right)^n}{n!} \nonumber \\
                                                                                                                                                                 &\leftstackrel{\frac{1}{r} \gg |\mathbf{k}| \sim \mZ}{\approx} \sum_{n=0}^{\infty} \, \tilde{\mu}^{2 \epsilon} \int \frac{d^{d-1} \mathbf{k} }{(2 \pi)^{d-1}}\,\frac{4 \pi \alpha}{\mathbf{k}^2 + \mZ^2}\,\frac{\left(i \mathbf{k} \cdot \mathbf{r}\right)^n}{n!} \nonumber \\
                                                                                                                                                                 &= \frac{\alpha}{r} \sinh(-\mZ r)\  \stackrel{r \to 0}{\longrightarrow} \  \frac{\alpha}{r}\left( -\mZ r - \frac{1}{6} \, \mZ^3 r^3 + \mathcal{O}(\mZ^5 r^5) \right)
\end{align}
which captures all screening contributions of the Yukawa potential, in particular the first subleading term for $r \to 0$. For the $T^3 Y$ linear combination at tree-level, this means
\begin{align}
  V^{T3Y}_{\rm tree}(r) = \frac{\alpha}{r} - \frac{\alpha}{r} \, e^{-\mZ r} \quad \stackrel{r\to 0}{\longrightarrow} \quad \alpha \mZ \, .
\end{align}
For the NLO correction, a similar cancellation happens, i.e., the leading pieces from $\mZ \ll |\mathbf{k}| \sim \frac{1}{r}$ cancel as SU(2) and U(1) disentangle at small-distance, respectively high energies. Therefore, the correction stems from the region $|\mathbf{k}|\sim \mZ \ll \frac{1}{r}$. However, an analytic extraction of the NLO coefficient is very challenging, as at NLO, the expanded Fourier transform of complicated one-loop structures, e.g., from the SM self-energies, needs to be evaluated. We can extract the light-fermionic contribution as all Fourier transforms from momentum to position space are analytically know, i.e., the full result can be Fourier transformed and then Taylor expanded around $r \to 0$. For electroweak and third-generation quarks, this is not possible as the Fourier transforms are not analytically known (cf. Appendix~B.2 of \cite{Beneke:2020vff}). The coefficients can be extracted numerically, or one could try to extract them by solving the integrals over the one-loop mass and momentum structures. Technically, the latter is equivalent to solving various two-loop integrals in $d=3$.

In any physical application, the coefficient is irrelevant for practical purposes, which is why we refrain from performing this analytic calculation. If $T^3Y$ is present for the potential, also $T^3 T^3$ and $YY$ are non-vanishing, which have an $1/r$ asymptotic behaviour at small distances. Already at $\mW r \approx 0.1$, both are more than two orders of magnitude larger. At even smaller distances, the differences grow linearly, making the constant term phenomenologically unimportant. For cases where different multiplets act as static sources and therefore $T^3 Y$ could appear without $T^3 T^3$ or $YY$, Higgs potentials with a tree-level term are expected, making the $T^3 Y$ UV limit again subleading.

The massive contributions are screened for large distances, and the leading correction comes from the corrections to photon exchange at tree-level. Split into the various gauge-invariant sub-parts, we find
\begin{align}
  \delta V^{T3Y}_{\rm light\, ferm.} (\mathbf{k}^2 \to 0) &=\frac{76}{9} \frac{\alpha^2}{\mathbf{k}^2} \ln \frac{\mathbf{k}^2}{\mZ^2} + \mathcal{O}(\mathbf{k}^0) \, , \\
    \delta V^{T3Y}_{\rm 3rd\, gen.\, quarks} (\mathbf{k}^2 \to 0) &=\frac{4}{9} \frac{\alpha^2}{\mathbf{k}^2} \ln \frac{\mathbf{k}^2}{\mZ^2} + \mathcal{O}(\mathbf{k}^0) \, , \\
    \delta V^{T3Y}_{\rm electroweak} (\mathbf{k}^2 \to 0) &= \frac{\alpha_2^2 s_W^2}{\mW^2} F(\mW,\mZ,\mh) \, .
\end{align}
The light fermionic and third-generation quark contribution -- namely there $b \bar{b}$-loops -- correct the large-distance behaviour logarithmically with a coefficient proportional to the electromagnetic beta function. The electroweak correction is given by the function $F(\mW,\mZ,\mh)$ in Appendix~\ref{app:asy} and is exponentially suppressed in position space. Nevertheless, the function provides a strong check on the calculation as it has to obey the screening theorem~\cite{Veltman:1976rt} for the Higgs mass dependence, which is indeed fulfilled. It evaluates to $-0.61209$ for on-shell parameters. Therefore the full correction to the $T^3 Y$ channel at large distances is given by
\begin{align}
  \delta V_{r \to \infty}^{T3Y}(r) = \frac{\alpha^2}{2 \pi r} \left(-\beta_{0,{\rm em}}\right) \left(\ln(\mZ r) + \gamma_E\right)
\end{align}
where $\beta_{0,{\rm em}} = -80/9$ is the electromagnetic beta function coefficient.

\subsubsection{$YY$ - linear combination}

\begin{figure}[t!] \centering
    \includegraphics[width=0.7\textwidth]{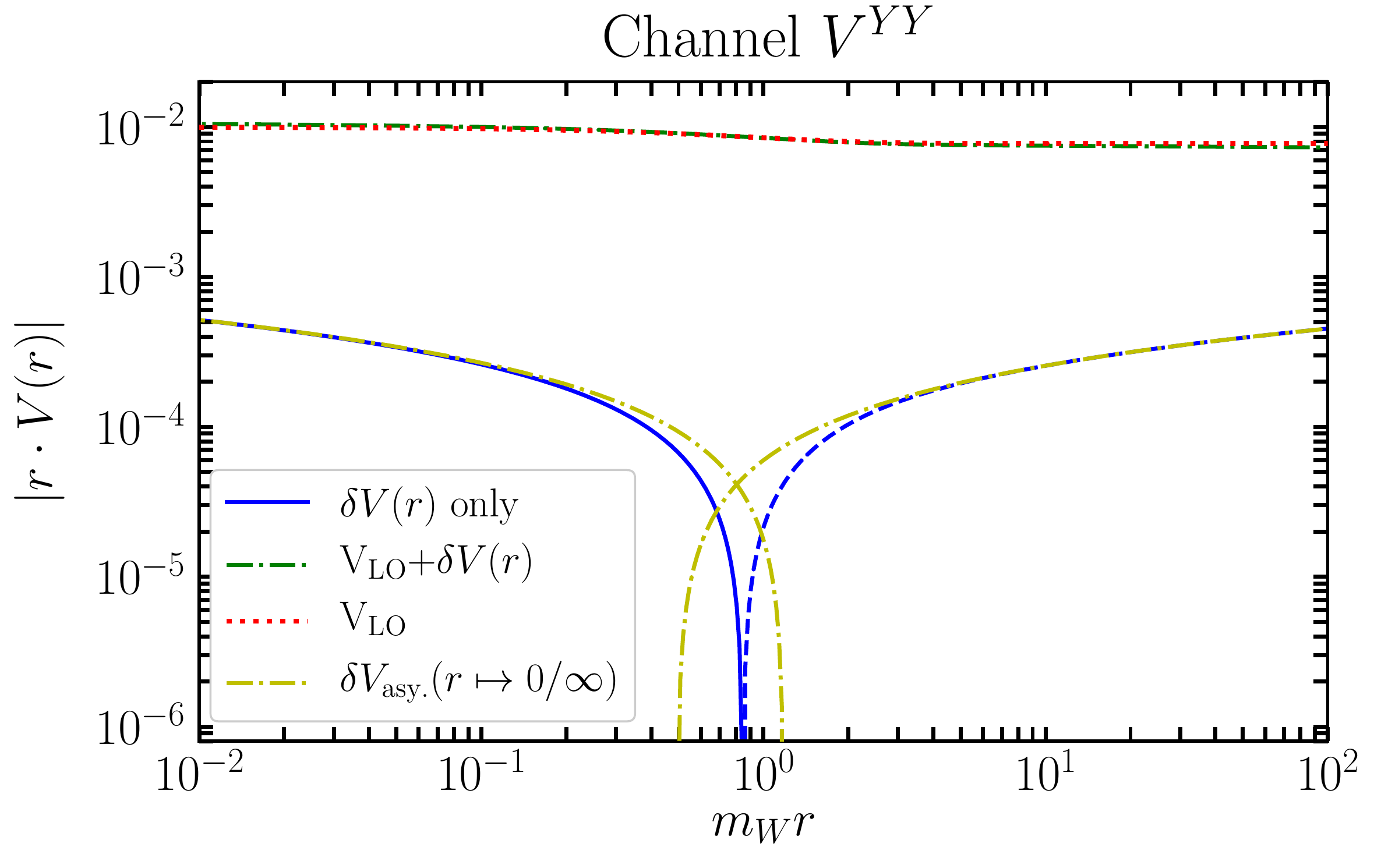}
    \hspace*{-0.725cm}\includegraphics[width=0.714\textwidth]{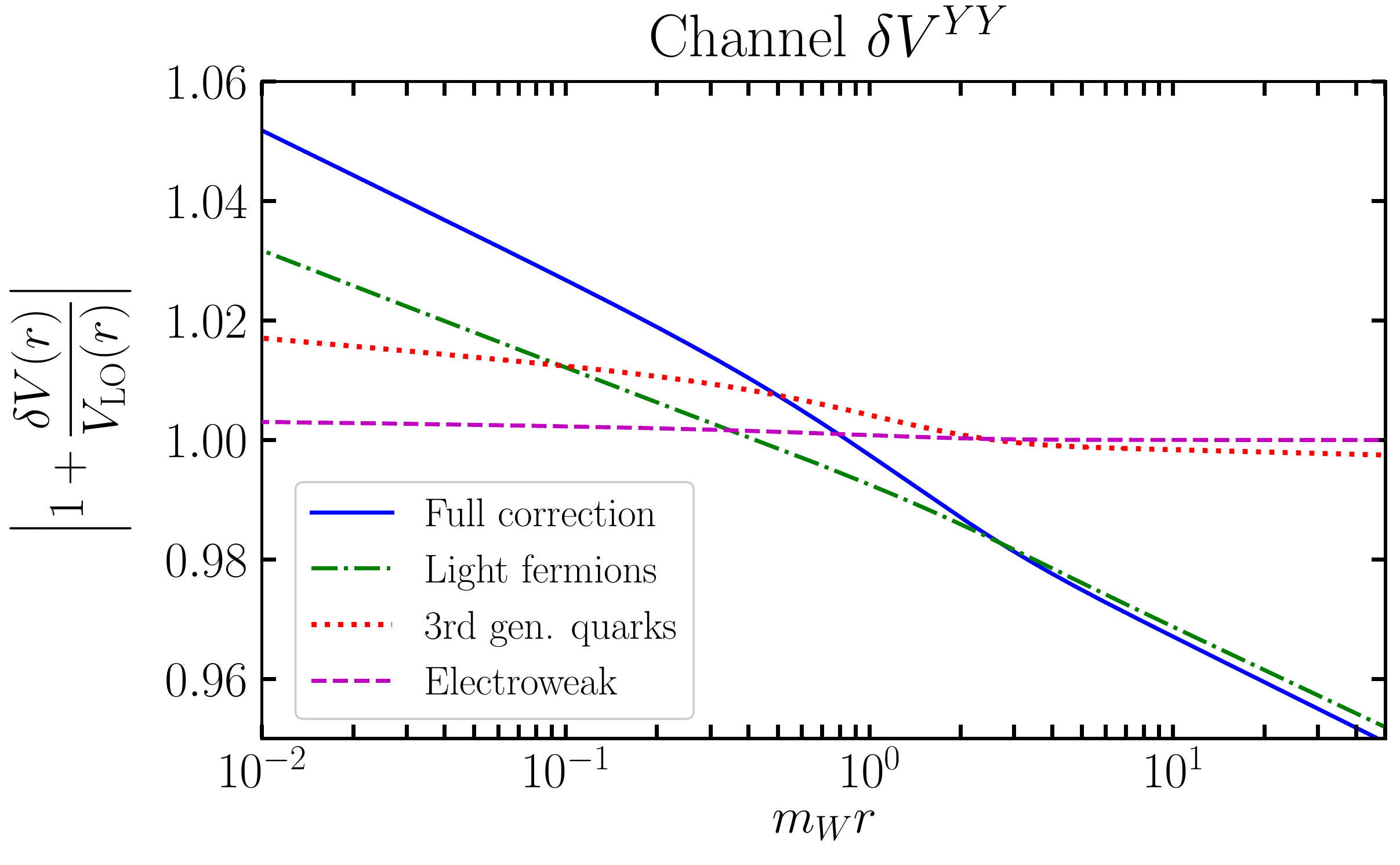}
    \caption{The NLO correction to the tree potential $V_{\rm tree}^{YY}$. The upper panel shows the modulus of the potential $|r \cdot V(r)|$ for the LO and NLO potential, the NLO contribution only, and the asymptotic behaviours (the change from solid to dashed indicates the sign change of the correction). In the lower panel, we show the ratio of the full NLO potential to the LO potential (blue solid) and separately for the three gauge invariant pieces identified in the text (other curves).} \label{fig:PotYY} 
  \end{figure}
In Figure~\ref{fig:PotYY}, the correction in the channel $\delta V^{YY}$ is shown. At small distances, as expected, we recover the potential correction to the unbroken U(1)$_Y$ theory, i.e., the various gauge-invariant sub-parts are corrected by beta function logarithms corresponding to their hypercharge
\begin{align}
  \delta V^{YY}_{\rm light\, ferm.} (\mathbf{k}^2 \to \infty) &= \frac{\alpha_1^2}{\mathbf{k}^2} \left(\frac{49}{9} \ln \frac{\mathbf{k}^2}{\mZ^2} + \frac{3 c_W^2}{s_W^2} \ln \frac{\mW^2}{\mZ^2}\right) \\
  \delta V^{YY}_{\rm 3rd\, gen.\, quarks} (\mathbf{k}^2 \to \infty) &= \frac{\alpha_1^2}{\mathbf{k}^2} \left(\frac{11}{9} \ln \frac{\mathbf{k}^2}{\mZ^2} + G(\mW,\mZ,\mt) \right)  \\
    \delta V^{YY}_{\rm electroweak} (\mathbf{k}^2 \to \infty) &=\frac{\alpha_1^2}{\mathbf{k}^2} \left(\frac{1}{6} \ln \frac{\mathbf{k}^2}{\mZ^2} + H(\mW,\mZ,\mh) \right) \, .
    \label{eq:YYasy}
\end{align}
The further functions $G(\mW,\mZ,\mt)$ and $H(\mW,\mZ,\mh)$ that correct the purely Coulombic piece for $r \to 0$ are given in Appendix~\ref{app:asy}. $G$ can be approximated to sub permille accuracy in the interval of $\pm 10\,{\rm GeV}$ around the on-shell top-mass $m_{t,{\rm os}} = 173.1\, {\rm GeV}$, by $G(\mW,\mZ,\mt) = 11.9389 + 4.40457 \cdot 10^{-4} \,{\rm GeV^{-2}} \times (m_t^2 - m_{t,{\rm os}}^2)$, for all other parameters at on-shell values. The function $H$ evaluates to $2.53229$ for on-shell parameters. To determine the full asymptotic behaviour in the $r \to 0$ limit, we define 
\begin{align}
  \Delta_2 =  \frac{3 c_W^2}{s_W^2} \ln \frac{\mW^2}{\mZ^2} + G(\mW,\mZ,\mt) + H(\mW,\mZ,\mh) \, ,
\end{align}
which in the $YY$-sector takes the form
\begin{align}
  \delta V^{YY}_{r \to 0} (r) &= \frac{\alpha_1^2}{2 \pi r} \left(- \beta_{0,Y} \left(\ln (\mZ r) +\gamma_E\right) + \frac{1}{ 2} \Delta_2 \right) \nonumber \\
  &\approx\frac{\alpha_1^2}{2 \pi r} \left(- \beta_{0,Y} \ln (\mZ r)  + 1.97247  \right)
  \label{eq:YYasy0}
\end{align}
where $\beta_{0,Y} = - \frac{49}{9} -  \frac{11}{9} - \frac{1}{6} = -\frac{41}{6}$, and the last line provides the numerical value for on-shell parameters. Of the numerical coefficient, the light-fermion term makes up $-4.46142$, the third-generation quarks $5.26395$, the electroweak terms $1.16994$ and the Euler-Mascheroni constant associated with the logarithm $-3.94431$.

Going to large distances, the leading correction, as for $T^3 Y$ above, stems from corrections to the photonic Coulomb potential. Therefore, the correction is again logarithmic in the electromagnetic beta function, and the electroweak part only gives a subleading contribution
\begin{align}
    \delta V^{YY}_{\rm light\, ferm.} (\mathbf{k}^2 \to 0) &=\frac{76}{9} \frac{\alpha^2}{\mathbf{k}^2} \ln \frac{\mathbf{k}^2}{\mZ^2} + \mathcal{O}(\mathbf{k}^0) \, , \\
    \delta V^{YY}_{\rm 3rd\, gen.\, quarks} (\mathbf{k}^2 \to 0) &=\frac{4}{9} \frac{\alpha^2}{\mathbf{k}^2} \ln \frac{\mathbf{k}^2}{\mZ^2} + \mathcal{O}(\mathbf{k}^0) \, , \\
    \delta V^{YY}_{\rm electroweak} (\mathbf{k}^2 \to 0) &=\frac{\alpha_2^2 s_W^2}{\mW^2} I(\mW,\mZ,\mh)
\end{align}
with the function $I(\mW,\mZ,\mh) = 0.75918$ for on-shell parameters, given in functional form in Appendix~\ref{app:asy} and obeying the screening theorem \cite{Veltman:1976rt}. Therefore, at large distances, we find
\begin{align}
  \delta V_{r \to \infty}^{YY}(r) = \frac{\alpha^2}{2 \pi r} \left(-\beta_{0,{\rm em}}\right) \left(\ln(\mZ r) + \gamma_E\right) \, .
  \label{eq:YYasyinf}
\end{align}
In contrast to the $T^3 T^3$ analogue, discussed on the wino channel $\chi^+ \chi^- \to \chi^+ \chi^-$ in \cite{Beneke:2020vff}, the beta function behaviour for large and small distances does not change sign for the $YY$ correction, as both limits correspond to an Abelian symmetry. In turn, this means that the correction changes sign in the most relevant region $\mW r \sim 1$, cf. Figure~\ref{fig:PotYY}.

\subsubsection{Special linear combinations of $T^3$ and $Y$}
\begin{figure}[t] \centering
    \includegraphics[width=0.8\textwidth]{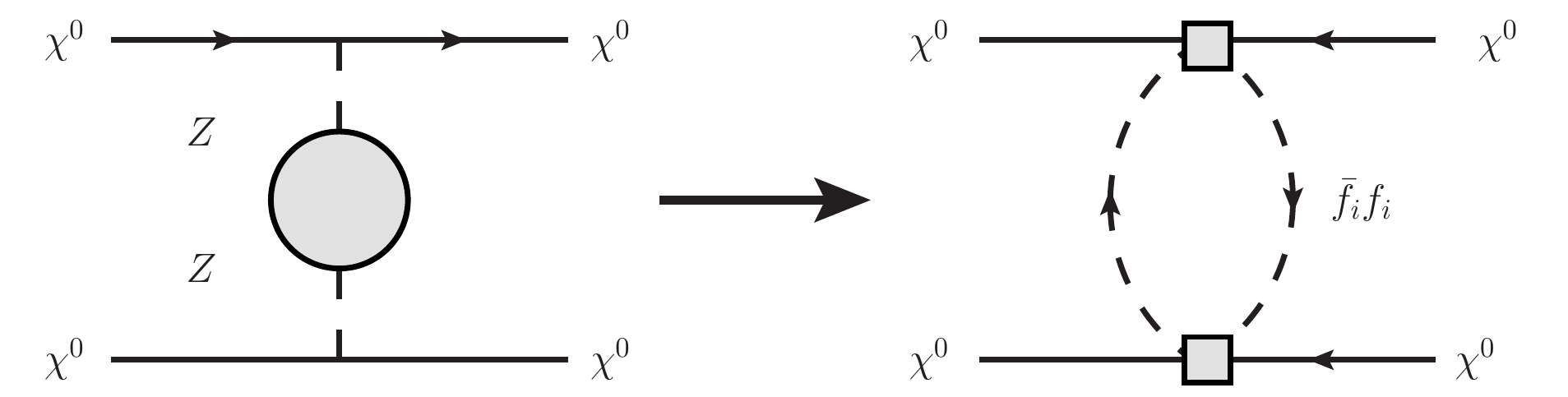}
  \caption{The relevant one-loop self-energy, that in the large distance, small momentum exchange region, gives the long-range asymptotic \eqref{eq:r5asymptoticZboson} for $\delta V^{T3T3} - 2\delta V^{T3Y} +\delta V^{YY}$.} \label{fig:Zonly} \end{figure}

\begin{figure}[t!] \centering
    \includegraphics[width=0.49\textwidth]{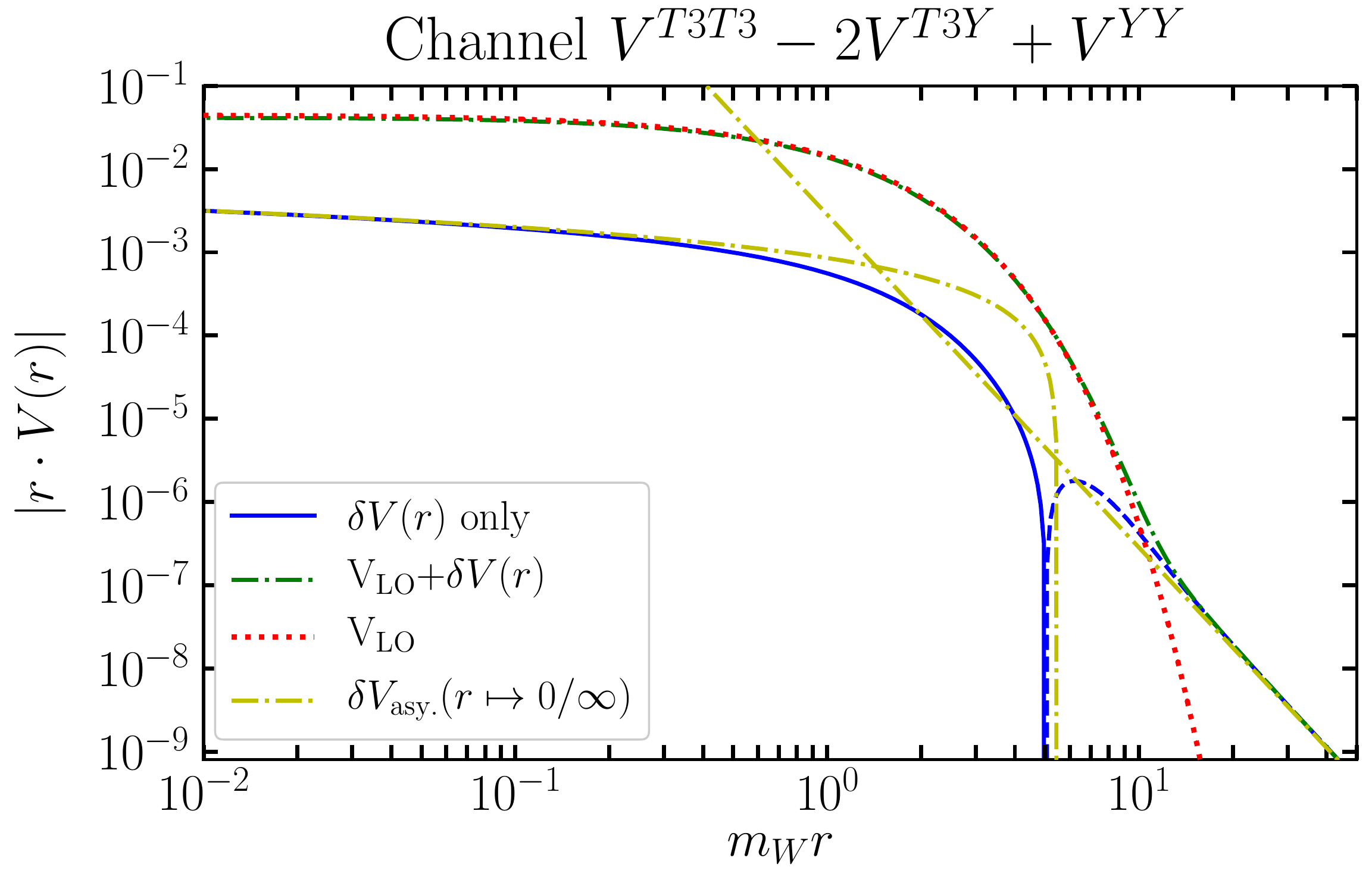}
    \hspace*{0.01\textwidth}\includegraphics[width=0.49\textwidth]{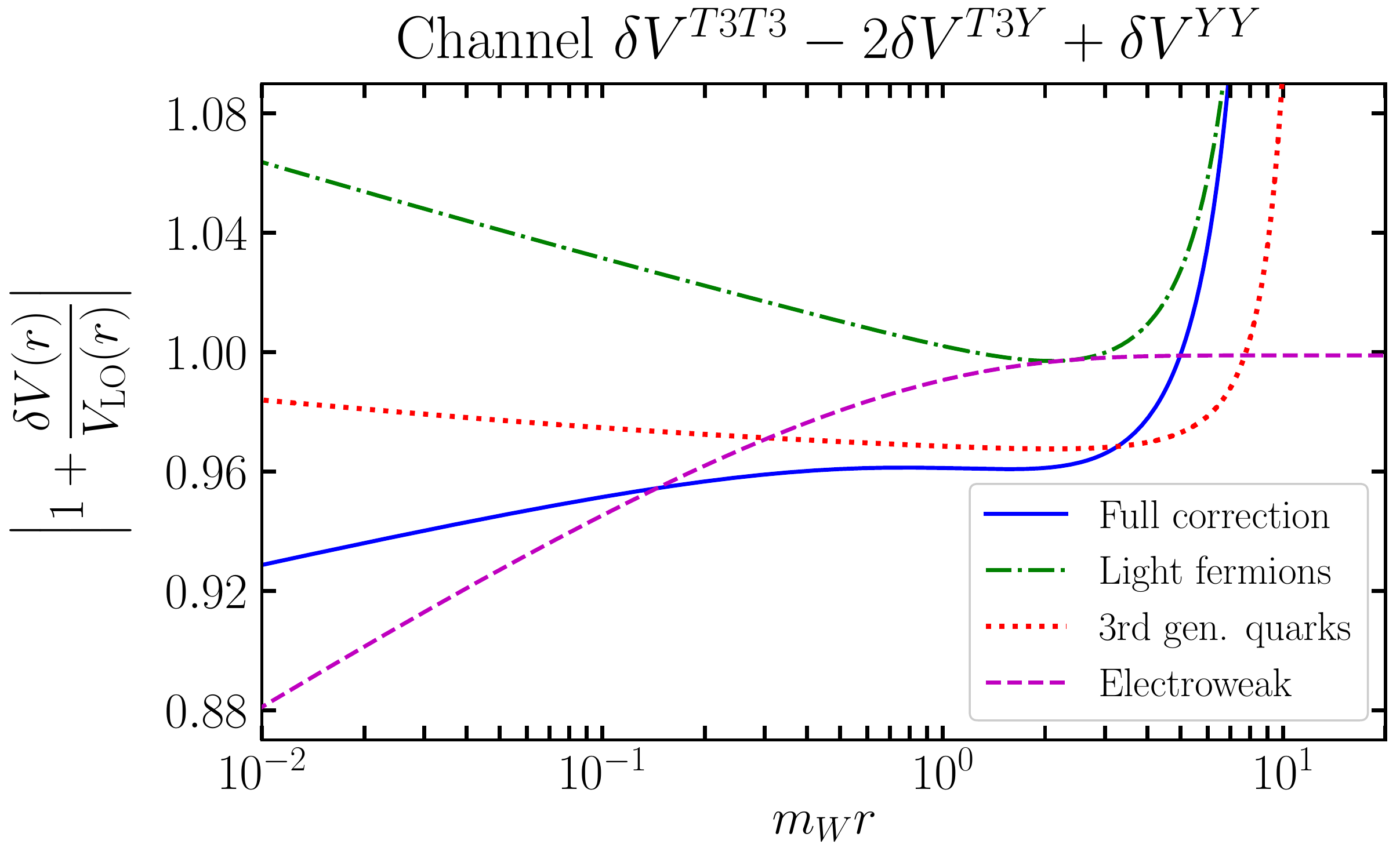}
    \includegraphics[width=0.49\textwidth]{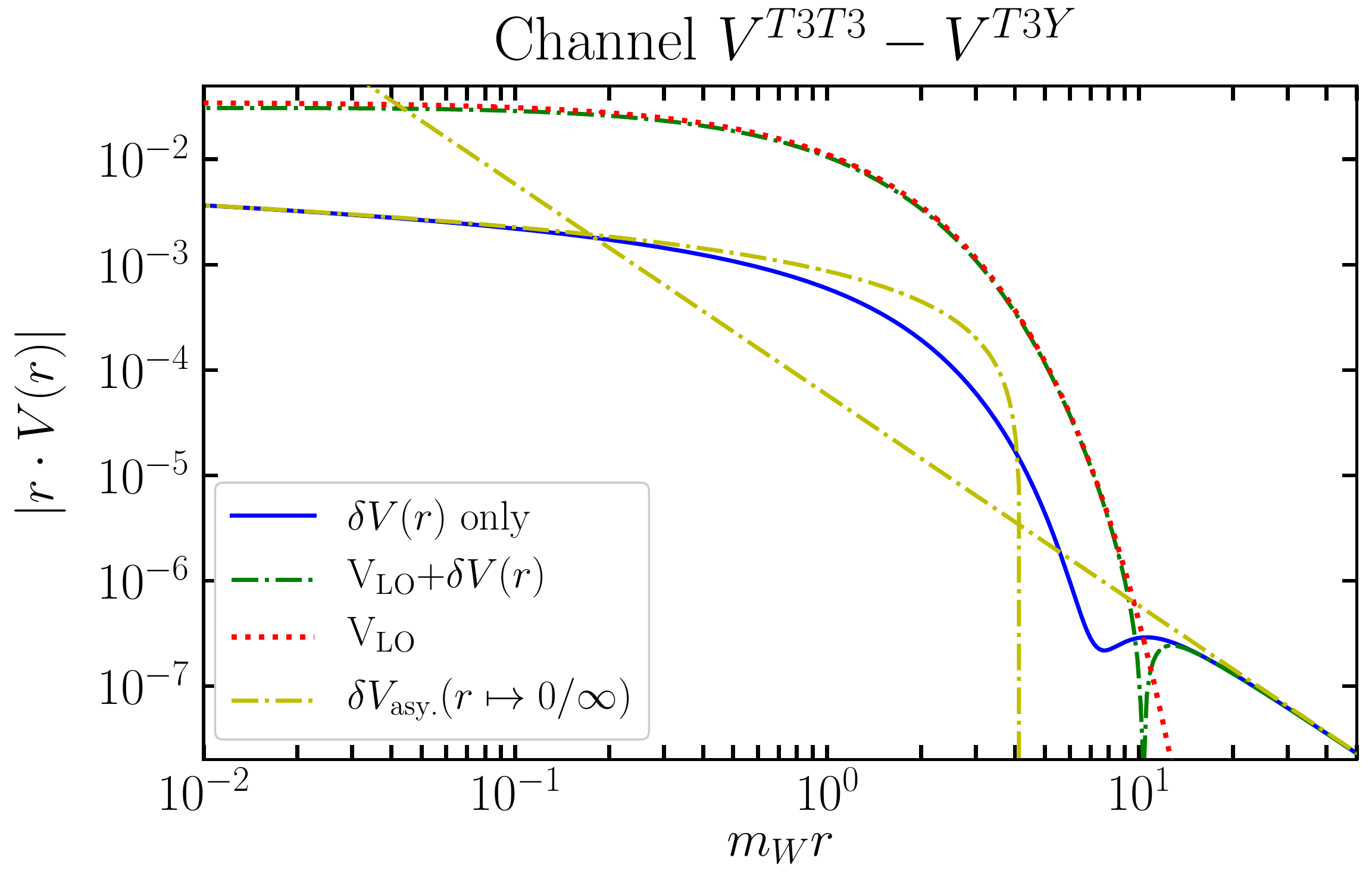}
    \hspace*{0.01\textwidth}\includegraphics[width=0.49\textwidth]{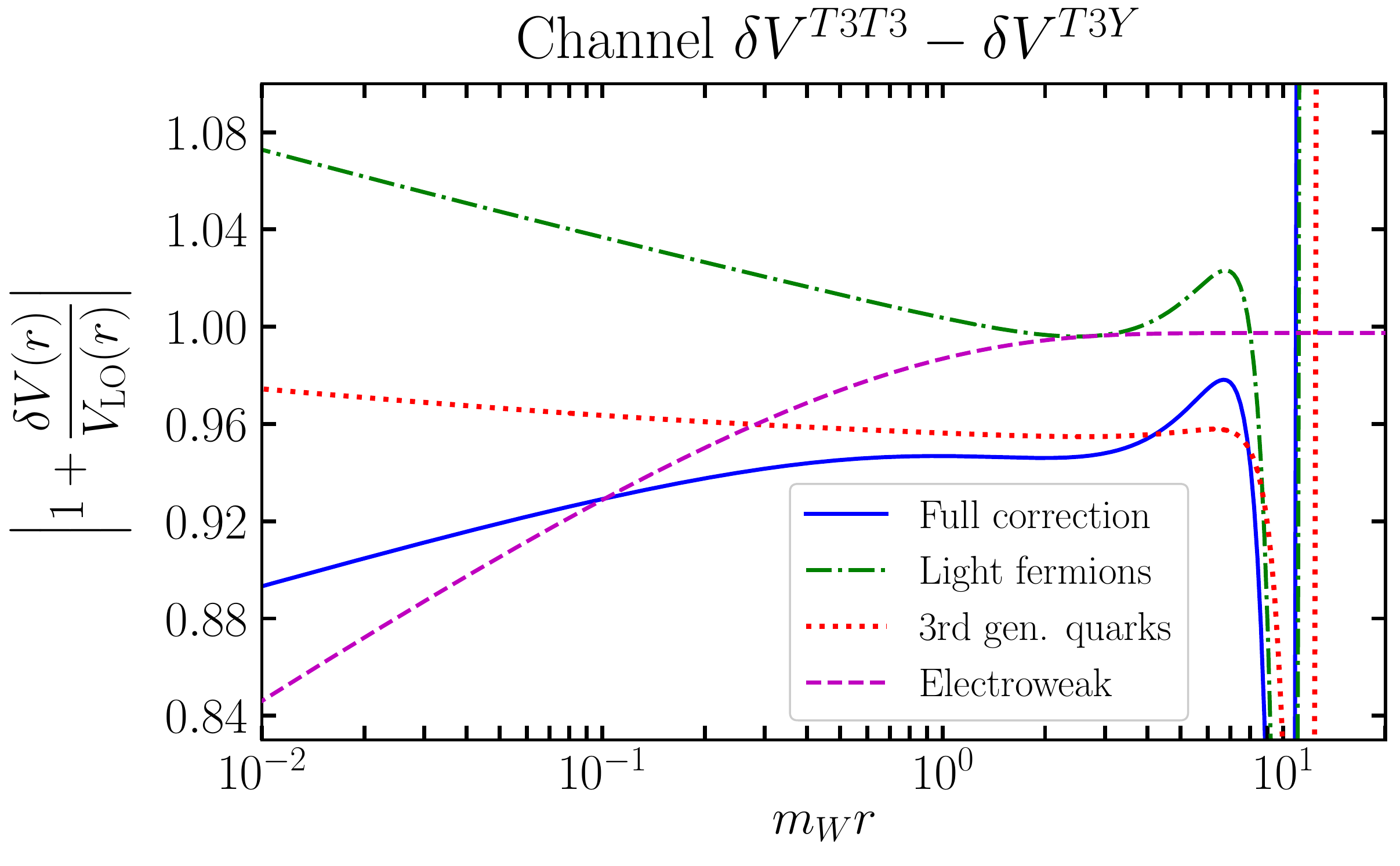}
    \includegraphics[width=0.49\textwidth]{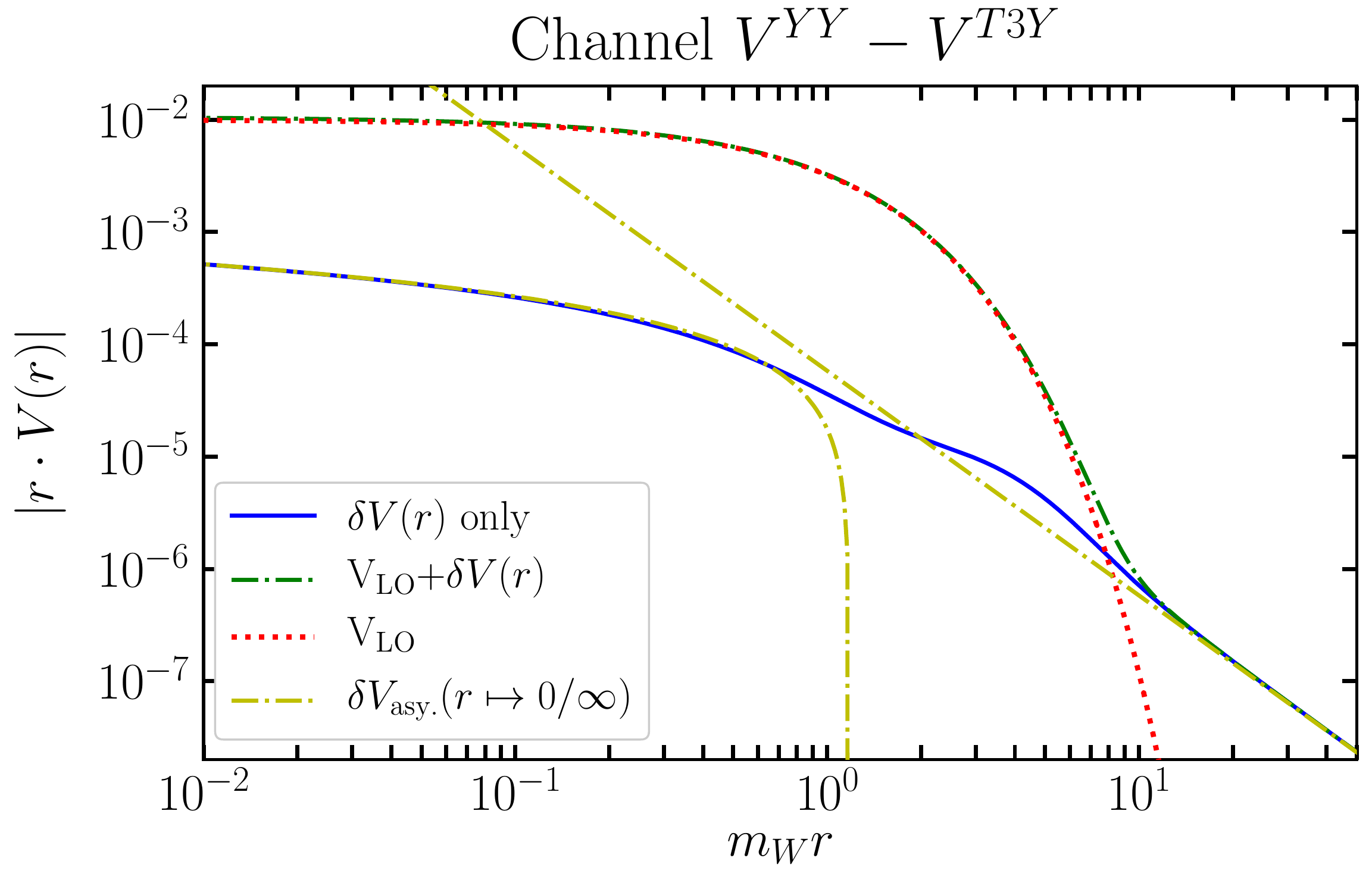}
    \hspace*{0.01\textwidth}\includegraphics[width=0.49\textwidth]{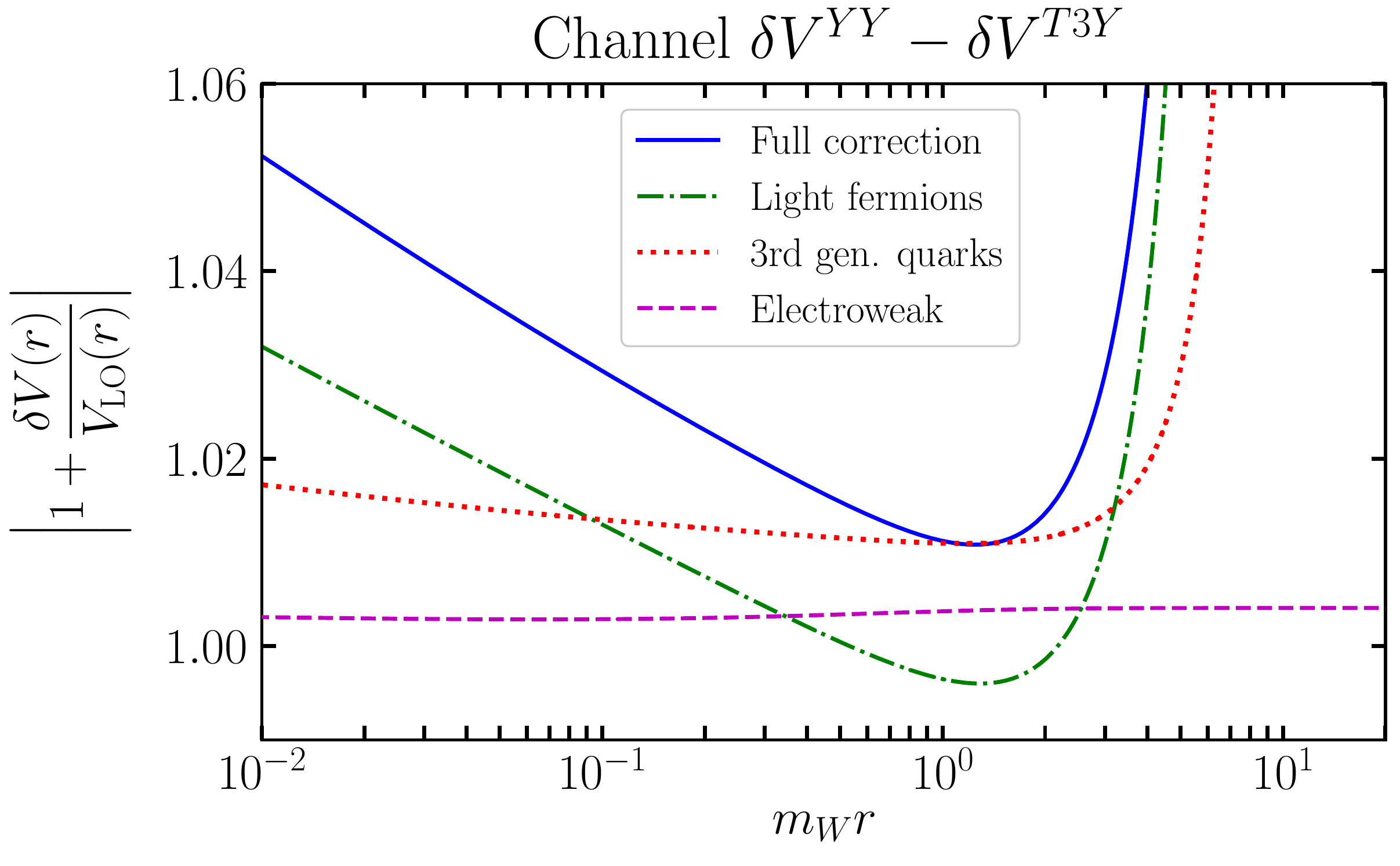}
    \caption{The NLO correction to the tree-level $Z$-boson only exchange potentials discussed in the text. The left panel shows the modulus of the potential $|r \cdot V(r)|$ for the LO and NLO potential, the NLO contribution only, and the large-distance asymptotic behaviour (the change from solid to dashed for the correction only line, indicates the sign change of this term). In the right panel, we show the ratio of the full NLO potential to the LO potential (blue solid) and separately for the three gauge invariant pieces identified in the text (other curves). The upper plots correspond to the linear combination given in \eqref{eq:lincomb_neutraldiag}, the middle and lower plots to the linear combinations in \eqref{eq:lincomb_chargeddiag}. }
\label{fig:PotZonly} 
  \end{figure}

For the $T^3 T^3$ projection, as well as the $T^3 Y$ and $YY$ cases discussed above, at large distances, the correction is dominated by the U(1)$_{\rm em}$ correction to the tree-level photonic Coulomb potential. In special cases, where the Coulomb potential is not present at tree-level, e.g., if $T^3_{R,ii} + Y_i = Q_i = 0$, the correction shows different IR behaviour, as only $Z$-exchange is possible at tree-level. Therefore, we have to reanalyze the large distance behaviour for the relevant linear combinations in these cases. 

We begin with the linear combination, for which both heavy particles are electrically neutral, i.e., $T^3_{ii} = -Y_i$ and $T^3_{jj} = -Y_j$. In this case, the relevant linear combination reads
\begin{align}
  \delta V^{T3T3 - 2 T3 Y + YY}_{Z-{\rm only}} = \delta V^{T3 T3} - 2 \delta V^{T3Y} + \delta V^{YY} \, .
  \label{eq:lincomb_neutraldiag}
\end{align}
At large distances, i.e. small momenta, massive propagators are not resolved anymore, which leads to the situation depicted in Figure~\ref{fig:Zonly}. The $Z$-boson propagators can be shrunk to a point, and the EFT is essentially a Four-Fermi theory. However, as the light fermions in the loop are massless from the perspective of the electroweak scale, they generate a long-range potential that on dimensional grounds has to scale as $r^{-5}$. We find the asymptotic behaviour
\begin{align}
  \delta V^{r \to \infty}_{Z-{\rm only}}(r) &=\frac{12 \alpha}{\pi s_W^2 c_W^2 \mZ^5 r^5} \left\{\frac{\alpha \mZ}{c_W^2 s_W^2}\left[\left(\frac{27 - 54 s_W^2 +76 s_W^4}{36}\right)+\left(\frac{9 - 12 s_W^2 + 8s_W^4}{72}\right)\right]\right\}\nonumber \\
                                            &=\frac{12 \alpha}{\pi s_W^2 c_W^2 \mZ^4 r^5} \frac{\Gamma_Z}{\mZ}
  \label{eq:r5asymptoticZboson}
\end{align}
where the $Z$-width is given by $m_Z \Gamma_Z = {\rm Im} \Sigma_T^{ZZ}(\mZ^2)$. In the first line, inside the curly brackets, the terms indicate the light fermionic and third-generation quark (i.e., $b\bar{b}$-loops) contributions to the $Z$-width, respectively. The behaviour is shown in Figure~\ref{fig:PotZonly}. The physics is completely analogous to the corresponding situation in the (off-diagonal) $W$-exchange channel (cf.~\cite{Beneke:2020vff}), and is known from the analogous situation of a long-range force due to massless neutrino exchange in atomic physics \cite{Feinberg:1968zz,Grifols:1996fk}. 

\begin{figure}[t] \centering
    \includegraphics[width=0.8\textwidth]{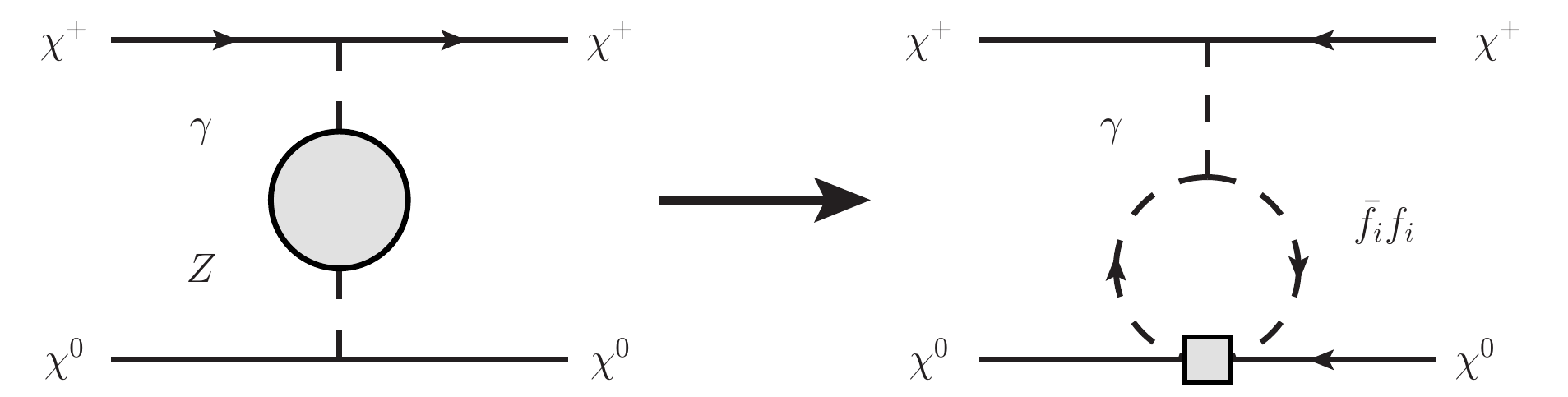}
  \caption{The relevant one-loop self-energy, that in the large distance, small momentum exchange region, gives the long-range asymptotic \eqref{eq:r3asymptotic} for $\delta V^{T3T3} - \delta V^{T3Y}$ and $\delta V^{YY} - \delta V^{T3Y}$.} \label{fig:T3T3minT3Yasy} \end{figure}

For an indirect detection analysis, e.g., of minimal DM, where the overall charge of the two-particle system is vanishing, the above is sufficient. In general, e.g., in relic abundance calculation or going away from DM, the restriction of both heavy particles being neutral for a $Z$-only Yukawa potential at tree-level is too restrictive. Without loss of generality, let us assume that the heavy particle $i$ is neutral $T^{3}_{R,ii} = -Y_i$ and the particle $j$ is electrically charged $T^3_{R,jj} \neq - Y_j$. The linear combination for the potential corrections is then
\begin{align}
  -Y_i T^3_{jj} \left(\delta V^{T3T3} - \delta V^{T3Y}\right) + Y_i Y_j \left(\delta V^{YY} - \delta V^{T3Y}\right)\, ,
  \label{eq:lincomb_chargeddiag}
\end{align}
meaning we have to examine $\delta V^{T3T3} - \delta V^{T3Y}$ and $\delta V^{YY} - \delta V^{T3Y}$. The asymptotic behaviour at large distances is now $r^{-3}$, as there is still the possibility of a photon coupling to the charged particle, as depicted in a concrete example in Figure~\ref{fig:T3T3minT3Yasy}. The correction is of opposite sign between the two linear combinations, in order to reproduce \eqref{eq:r5asymptoticZboson} if the two asymptotics are added
\begin{align}
  \left(\delta V^{T3T3} - \delta V^{T3Y}\right)( r \to 0) &= -\left(\delta V^{YY} - \delta V^{T3Y}\right)( r \to 0) \nonumber \\
                                                          &=-\frac{\alpha_1 \alpha_2}{18 \pi \mZ^2 r^3} \left\{\left(38 c_W^2 - 38 s_W^2 -11\right) +\left(2 c_W^2 - 2 s_W^2 +1\right) \right\}
  \label{eq:r3asymptotic}
\end{align}
where the terms in the curly brackets correspond to light fermions and third-generation quarks, respectively. As indicated in Figure~\ref{fig:T3T3minT3Yasy}, the terms arise from massless fermion loops in $\Sigma_T^{\gamma Z}$, and if added the middle and lower panel of Figure~\ref{fig:PotZonly} -- adjusted for the sign -- indeed reproduce the upper panel.

\subsection{Fitting functions for the NLO correction}
To make the result easily usable, we provide fitting functions for the various channels using the on-shell parameters discussed above. The fitting functions approximate the full numerical Fourier transform of the momentum space potentials in Section~\ref{sec:NLOpotentials} to permille level accuracy for the region of interest in practical calculations, e.g., of the Sommerfeld enhancement. In order to shorten notation, we introduce the variables $x = \mW r$, and $L = \ln x$. Note that for the asymptotic regions $x \gtrsim 10^2$ and $x \lesssim 10^{-2}$, the asymptotic behaviours discussed above may be used. Differences between asymptotic behaviours and full Fourier transform are permille level or even below in these regions.

\paragraph{Fitting function for off-diagonal $W$-exchange}
A piecewise function efficiently fits the off-diagonal $W$-exchange contribution due to a sign change of the correction around $x_0 = 555/94$, and the different functional forms of the long and short-distance behaviour. The function is introduced in \cite{Beneke:2019qaa} for the wino in the channel $\chi^0 \chi^0 \to \chi^+ \chi^-$, which amounts to an overall minus sign from wino coupling factors. We repeat the function for completeness 
\begin{align}
  \delta &V_{\text{fit}}^{W}(r) = \frac{2595 \alpha^2_2}{\pi r} \times \left\{\begin{array}{r}
        - \text{exp} \left[ -\frac{79 \left(L-\frac{787}{12}\right)
     \left(L-\frac{736}{373}\right)\left(L-\frac{116}{65}\right)
     \left(L^2-\frac{286L}{59}+\frac{533}{77}\right)}{34
     \left(L-\frac{512}{19}\right)\left(L-\frac{339}{176}\right)
     \left(L-\frac{501}{281}\right)\left(L^2-\frac{268
     L}{61}+\frac{38}{7}\right)}\right] ,\quad  x<x_0 \\[0.5cm]
 \text{exp}\left[-\frac{13267 \left(L-\frac{76}{43}\right)
 \left(L-\frac{28}{17}\right)\left(L+\frac{37}{30}\right)
 \left(L^2-\frac{389L}{88}+\frac{676}{129}\right)}{5
 \left(L-\frac{191}{108}\right)\left(L-\frac{256}{153}\right)
 \left(L+\frac{8412}{13}\right) \left(L^2-\frac{457
 L}{103}+\frac{773}{146}\right)}\right] ,\quad  x>x_0 \end{array}\right. \, ,
      \end{align}
      to be used in \eqref{eq:CasimirWexchange}
\paragraph{Fitting function for $T^3 T^3$}
For the projection on the $T^3 T^3$ component of the photon and $Z$ potential, the fitting function 
      \begin{align}
        \delta &V_{\text{fit}}^{T3T3} = \frac{\delta V_{r \to \infty}^{T3T3} }{1+ \frac{32}{11 }x^{-\frac{22}{9}}} + \frac{\delta V_{r \to
0}^{T3T3}}{1+\frac{7}{59} x^{\frac{61}{29}}} -
\frac{\alpha}{r} \left[\frac{-\frac{1}{30} + \frac{4}{135} \ln x}{1 +
\frac{58}{79} x^{-\frac{17}{15}}+\frac{1}{30} x^{\frac{119}{120}} +
\frac{8}{177} x^{\frac{17}{8}}}\right]\,,
\end{align}
is introduced and can be used in \eqref{eq:CasimirgamZ}. The function was first given -- with adjusted coupling factors -- for the corresponding wino channel in \cite{Beneke:2019qaa}. The asymptotic behaviours for $r \to 0,\infty$ are given by (with rationalized coefficients)
\begin{align}
  \delta V_{r \to 0}^{T3T3}(r) &= \frac{\alpha_2^2}{2 \pi r} \left(\beta_{0,{\rm SU(2)}} \ln (\mZ\,  r) - \frac{1960}{433} \right) \label{eq:rto0T3T3} \\
\delta V_{r \to \infty}^{T3T3}(r) &= \frac{\alpha^2}{2 \pi r} \,\beta_{0,{\rm em}} \,(\ln (m_Z r)+\gamma_E) \label{eq:rtoinfT3T3}
\end{align}
with $\beta_{0,{\rm em}} = -80/9$ and $\gamma_E$ the Euler-Mascheroni constant. For an in detail discussion on the asymptotic behaviours and the full functional form in terms of $\mW,\mZ,\mh,\mt$, see \cite{Beneke:2020vff} where these are discussed in great detail for the wino channel $\chi^+ \chi^- \to \chi^+ \chi^-$.
\paragraph{Fitting function for $YY$}
For the pure hypercharge projection $YY$, we again fit with a piecewise function, as there is a sign change around $x_1=\frac{718}{853}$. The function to be used in \eqref{eq:CasimirgamZ} is given by
\begin{align}
  \delta V^{YY}_{\rm fit}(r) &= \left\{ \begin{array}{cc}
\displaystyle \frac{\delta V^{YY}_{r \to 0} }{1+ \frac{97}{12306436} e^{-\frac{484}{119} x}} - \frac{\alpha^2}{r} \frac{\frac{71}{104} + \frac{145}{2109} x}{1+ \frac{262}{185}e^{-\frac{961}{412}x}} + \frac{\frac{41}{19 r } + \frac{42}{11} \mW}{1+126870 \,e^{\frac{484}{119}x}} \,, &  x\leq x_1 \\[0.6cm]
\displaystyle \frac{\delta V^{YY}_{r \to \infty}}{1+e^{\frac{356}{185} - \frac{191}{179} x}} + \frac{\alpha^2}{r} \left[\frac{-\frac{85336}{177} - \frac{272 x}{21}}{1+ e^{\frac{398}{373} x- \frac{127}{66}}} + \frac{\frac{4402}{9} - \frac{959 \sqrt{x}}{93} + \frac{564 x}{35}}{1+e^{\frac{643}{602}x- \frac{2157}{1120} }}\right] \,, &  x > x_1 \, 
  \end{array}\right. \, .
\end{align}
The asymptotic behaviours in the $r \to 0, \infty$ limit were given above in \eqref{eq:YYasy0} and \eqref{eq:YYasyinf}.
\paragraph{Fitting function for $T^3 T^3 + T^3 Y$}
Separately fitting the channel $T^3 Y$ to sufficient accuracy proves challenging. As $T^3 T^3$ and $YY$ are always non-zero as long as both heavy sources are within the same multiplet, we fit $T^3 T^3 + T^3 Y$, which provides a much better fit quality compared to only fitting $T^3 Y$. Together with $T^3 T^3$ and $YY$, the fitting functions still cover all relevant linear combinations, except those with special IR dynamics below. The resulting fitting function to be used within \eqref{eq:CasimirgamZ} is given by
\begin{align}
 \left( \delta V^{T3T3} + \delta V^{T3Y} \right)_{\rm fit}\!(r) = \frac{\delta V^{T3T3}_{r \to 0}}{1+ \frac{62}{161} x^{\frac{203}{185}}}\! +\! \frac{2\, \delta V^{T3T3}_{r \to \infty}}{1+ \frac{807}{280} x^{-\frac{267}{268}}}\! + \! \frac{\frac{\alpha}{r} \left(- \frac{77}{57} + \frac{71}{485} L - \frac{64}{263} L^2\right)}{\left(1+\frac{268}{7} x^{-\frac{220}{109}}\right)\left(1+ \frac{686}{145} x^{\frac{48}{43}}\right)}\,,
\end{align}
where the necessary asymptotic behaviours are given in \eqref{eq:rto0T3T3} and \eqref{eq:rtoinfT3T3}.

\paragraph{Fitting function for $T^3 T^3 - 2 T^3 Y + YY$}
As discussed above, if the tree-level photon exchange vanishes even though $Z$-exchange is still possible, special IR behaviour of the potentials is observed. If the $Z$-only potential from the linear combination $T^3 T^3 -2 T^3 Y + YY$ appears in \eqref{eq:CasimirgamZ}, we use a piecewise fitting function, due to the sign change in the correction around $x_2 = \frac{1382}{275}$
\begin{align}
& \left( \delta V^{T3T3} \right.  -2\delta V^{T3 Y} \left.+\delta V^{YY} \right)_{\rm fit}(r) \nonumber \\
&= \frac{\alpha^2}{r} \left\{ \begin{array}{cc}
\! \! \! - \exp \left[\frac{597 \left(L - \frac{1667}{108}\right) \left(L- \frac{837}{503}\right) \left(L - \frac{55}{34}\right) \left(L-\frac{961}{911}\right) \left(L^2 - \frac{1715}{563}L + \frac{2072}{587}\right)}{83\left(L - \frac{841}{31}\right) \left(L - \frac{292}{133}\right) \left(L - \frac{1733}{1054} \right) \left(L - \frac{2105}{1302}\right) \left(L^2 - \frac{729}{263} L + \frac{1422}{451}\right)}\right] \, ,& x \leq  x_2 \\[0.6cm]
\exp \left[-\frac{3710 \left(L - \frac{637}{396}\right) \left(L + \frac{2999}{875}\right) \left(L^2 - \frac{7506}{1819} L + \frac{1097}{240}\right)\left(L^2 - \frac{5789}{2180} L + \frac{2399}{1326}\right)}{933 \left(L- \frac{398}{247}\right) \left(L- \frac{933}{613}\right) \left(L + \frac{2137}{662}\right)  \left(L^2 - \frac{2785}{664} L + \frac{8311}{1749}\right)}\right]\, , & x > x_2 
\end{array} \right. \, .
 \label{eq:Zonly_fitf}
\end{align}

\paragraph{Fitting function for $T^3 T^3 - T^3 Y$}
Finally, we need to cover the two cases with $r^{-3}$ long-distance behaviours. For $T^3 T^3 - T^3 Y$ in \eqref{eq:CasimirgamZ}, we fit with a piecewise function
\begin{align}
  &\left(\delta V^{T3T3}- \delta V^{T3Y}\right)_{\rm fit} \nonumber \\
  &\quad= - \frac{\alpha^2}{r} \left\{ \begin{array}{cc}
  \exp \left[\frac{1823 \left(L - \frac{262}{15}\right) \left(L - \frac{100}{91}\right) \left(L^2 - \frac{641}{145} L + \frac{1226}{177}\right) \left(L^2 - \frac{1434}{341} L + \frac{4709}{1052}\right)}{243 \left(L - \frac{1549}{52}\right) \left( L - \frac{925}{319} \right) \left(L^2 - \frac{1438}{351} L + \frac{642}{151}\right) \left(L^2 - \frac{595}{158} L + \frac{1491}{281}\right)} \right]\,, & x \leq x_3 \\[0.6cm]
  \exp \left[- \frac{902 \left(L - \frac{1229}{169}\right) \left(L - \frac{29}{279}\right) \left(L^2 - \frac{1331}{278} L + \frac{327}{56}\right) \left(L^2 - \frac{2920}{747} L + \frac{1013}{258}\right)}{\left(L - \frac{3891}{535}\right) \left(L + \frac{5687}{13}\right) \left(L^2 - \frac{627}{131} L +\frac{391}{67}\right) \left(L^2 - \frac{811}{207} L +\frac{661}{169}\right)}\right]\,, & x > x_3
  \end{array} \right. \, ,
\end{align}
with $x_3 = \frac{950}{119}$.

\paragraph{Fitting function for $YY - T^3 Y$}
For the linear combination $YY-T^3 Y$ in \eqref{eq:CasimirgamZ}, we find
\begin{align}
&\left(\delta V^{YY} - \delta V^{T3 Y}\right)_{\rm fit} \nonumber \\
&\quad =\frac{\alpha^2}{r} \left\{ \begin{array}{cc} 
\exp \left[\frac{518 \left(L -\frac{496}{19}\right) \left(L + \frac{106}{283}\right) \left(L^2 - \frac{392}{185} L + \frac{827}{394}\right) \left(L^2 + \frac{67}{189} L + \frac{1473}{158}\right)}{83 \left(L - \frac{395}{8}\right) \left(L - \frac{747}{236}\right) \left(L^2 - \frac{659}{445} L + \frac{193}{108} \right) \left(L^2 + \frac{347}{2171} L + \frac{677}{81}\right)}\right]\, , & x \leq x_4\\[0.6cm]
\exp \left[ \frac{1636 \left(L - \frac{837}{79}\right)\left(L + \frac{147}{397}\right) \left(L^2 - \frac{1073}{265} L + \frac{801}{173}\right) \left(L^2 - \frac{1805}{727} L + \frac{2630}{1127}\right)}{11 \left(L - \frac{687}{64}\right) \left(L - \frac{1981}{24}\right) \left(L^2 - \frac{2457}{601} L + \frac{520}{111}\right) \left(L^2 - \frac{449}{179} L + \frac{312}{115}\right)}\right]\, , & x > x_4
  \end{array} \right. \, ,
  \label{eq:fit_YYT3Y}
\end{align}
with $x_4 =\frac{379}{189} $.

\subsubsection{Accuracy of the fitting functions}
The accuracy for the off-diagonal $W$-exchange and the diagonal $T^3 T^3$-exchange fitting functions has already been discussed in depth in \cite{Beneke:2019qaa}. It was found that the fitting functions provide permille level accurate results, except in the absolute vicinity of a sign change ($W$-exchange). For practical calculations, e.g., of the Sommerfeld factors in indirect detection, this translates into subpermille accurate results, except close to zero-energy resonance positions where deviations of up to three permille are found.

\begin{figure}[t]
\centering
\includegraphics[width=0.7\textwidth]{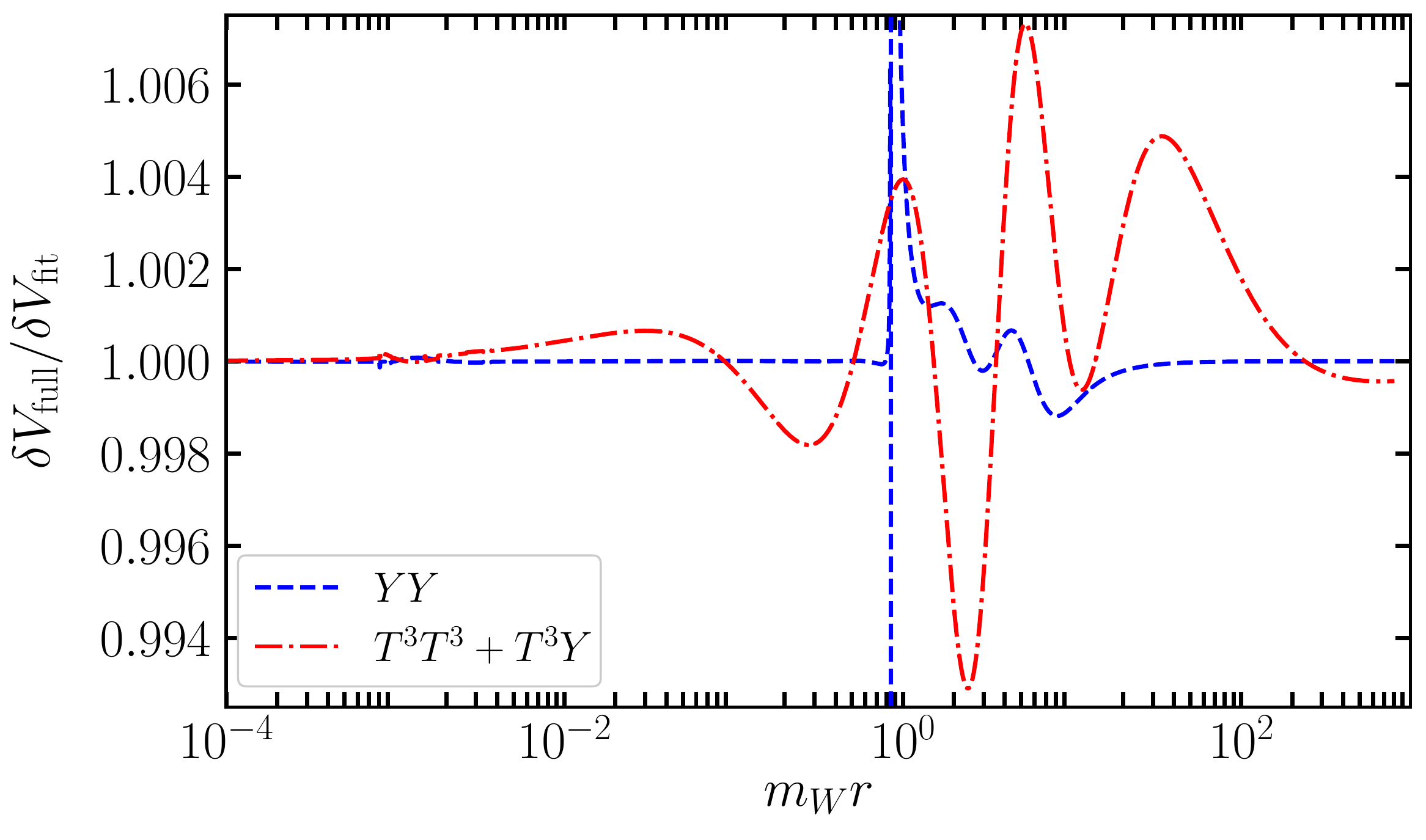}
\includegraphics[width=0.7\textwidth]{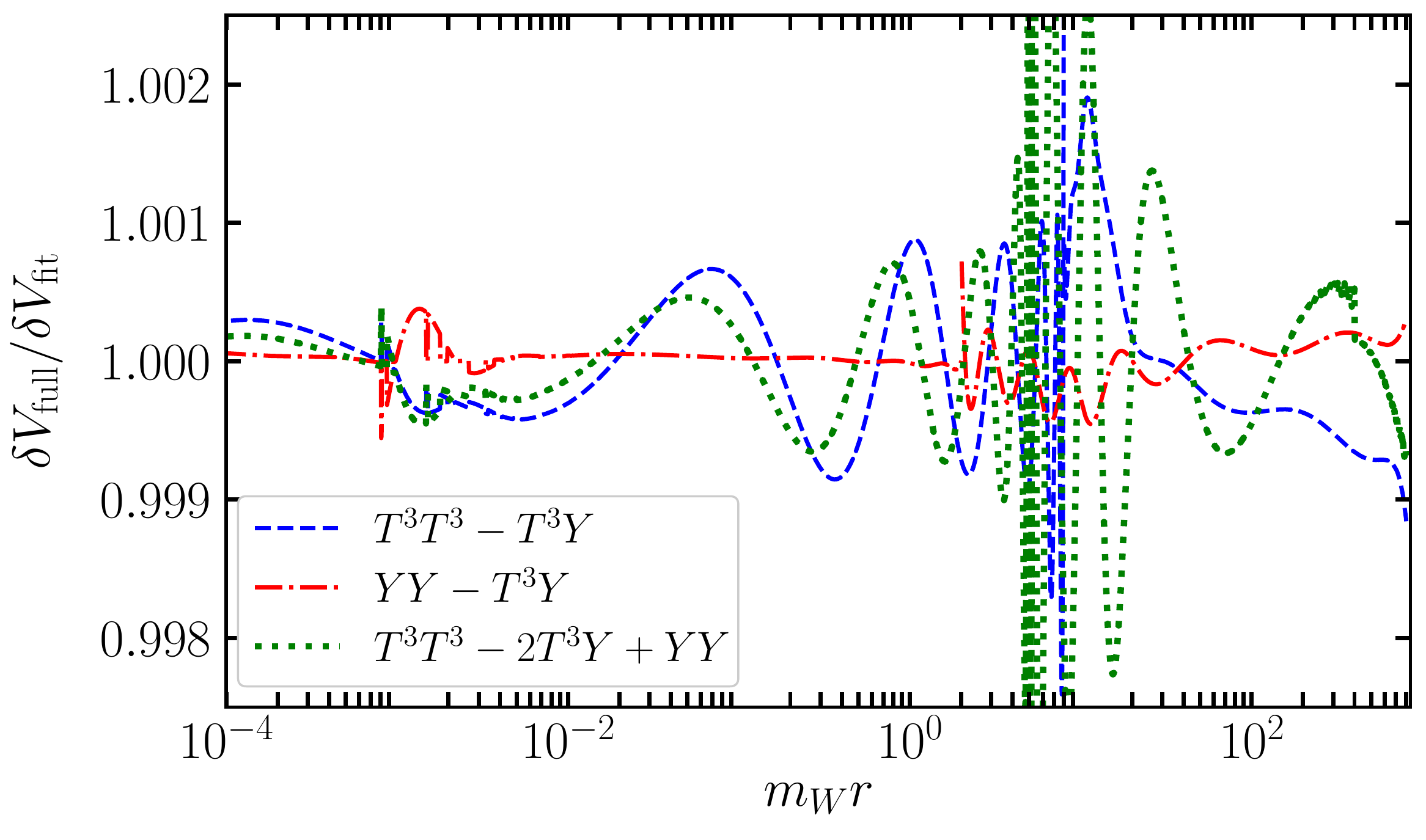}
\caption{Ratio of the numerical Fourier transform of the potential
    correction to the fitting functions in the variable $ x = m_W r$ for the
channels $YY$ (blue/dashed) and $T^3T^3 + T^3 Y$ (red/dash-dotted) in the upper panel. In the lower panel, the same ratio is shown for the channels with power-like large distance asymptotic behaviours $T^3 T^3 - T^3 Y$ (blue/dashed), $YY - T^3 Y$ (red/dash-dotted) and $T^3 T^3 - 2 T^3 Y + YY$ (green/dotted).}
\label{fig:qualityfitf}
\end{figure}

The comparison of the fitting functions against the full numerical solution of the Fourier transformation from position to momentum space in the new channels is shown in Figure~\ref{fig:qualityfitf}. In the upper panel, the channels $YY$ (blue/dashed) and $T^3 T^3 + T^3 Y$ (red/dash-dotted) are shown both agreeing on the permille level with the numerical solution, except very close to the sign change in $YY$, where the ratio of full numerical potential to fitting function naturally blows up.

We also find permille level accuracy for cases with special long-distance behaviours (lower panel), except at the sign changes. For practical purposes, the results are sufficient to allow an accurate solution of the Schr\"odinger equation. To validate this statement, we use the higgsino model implemented as discussed in \cite{Beneke:2019gtg}. We find that the accuracy is similar to the wino case \cite{Beneke:2019qaa} with a maximum of two permille deviation at the exact resonance positions and even subpermille accuracies for the Sommerfeld factors off resonance. Implicitly, this tests all fitting functions, except $T^3 T^3 - T^3 Y$ and $YY - T^3 Y$, which would, e.g., appear in the relic abundance calculation. Given the precise agreement of numerical Fourier transform and fitting function shown in Figure~\ref{fig:qualityfitf}, we expect similar accuracy if these fitting functions are used to solve the Schr\"odinger equation.

\subsection{General comments}
Before concluding, let us make a few remarks on renormalization schemes and parameter dependencies without repeating a similar in-depth discussion as provided for the pure SU(2) part in the wino case \cite{Beneke:2020vff}. At small distances, the beta function logarithms for $r \to 0$ cause a breakdown of perturbation theory in the channel $\delta V^{YY}$ (the channel $\delta V^{T3Y}$ does not suffer this divergence, as SU(2) and U(1) disentangle for high-energies). The dangerous logarithmic behaviour is cured by a renormalization scheme conversion to the $\overline{\rm MS}$-scheme that absorbs the leading logarithms. For practical calculations, e.g., of the Sommerfeld enhancement, where the region $\mW r \sim 1$ dominates, the logarithmic behaviour of the on-shell result does not change the result by much in comparison to the $\overline{\rm MS}$ treatment. In the infrared, the $\overline{\rm MS}$-scheme suffers problems, e.g., in the off-diagonal $W$-exchange channel with $r^{-5}$ long-range behaviour, an unphysical $r^{-3}$ tail is induced in $\overline{\rm MS}$. Therefore, it is not clear which scheme is superior for practical calculations \cite{Beneke:2020vff}. The differences for final results are compatible with the expected differences between two renormalization schemes at one-loop.

Similarly, the parameter dependence for the new potentials involving hypercharge is similar to the pure SU(2) case \cite{Beneke:2020vff}. The only input parameter with a significant dependence and a sizeable modification of the one-loop potentials is the top quark mass as exemplified for the function $G$ below Eq.~\eqref{eq:YYasy}. For the mixed $T^3 Y$ component, a similar relative dependence is expected. However, as the contribution is subdominant for $r \to 0$ compared to $T^3 T^3$ and $YY$, a further investigation is not warranted. Furthermore, as the massive self-energy loops decouple in the infrared, the top mass dependence is also unimportant at large distances. Therefore, from a phenomenological standpoint, the top mass dependence to $T^3 Y$ is irrelevant.

\section{Conclusion}
\label{sec:conclusions}
In this paper, we have shown that the NLO correction to the electroweak potentials induced by $W,Z$, and photon exchange is universal and follows a ``Casimir-like" scaling. Thereby, we have unravelled a ``low-energy" property of the SM gauge-bosons. Previously, the NLO correction was discussed for wino DM \cite{Beneke:2020vff}, which covers the cases of pure SU(2) interactions ($T^+ T^-$ and $T^3 T^3$). In addition, we also calculated the SU(2) and hypercharge mixing contribution ($T^3 Y$) and the pure hypercharge correction ($YY$) for the SM.

For the general case of arbitrary hypercharge, we also identify linear combinations with unique infrared behaviour, namely when the electric charge of the heavy particles vanishes. For the case of $T^3 T^3 - 2 T^3 Y + YY$, we find an $r^{-5}$ long-distance tail from light-fermion loops in pure $Z$-exchange known in the neutrino context \cite{Feinberg:1968zz,Grifols:1996fk} and first discussed in the DM context in \cite{Beneke:2019qaa} (for the case of $W$-exchange). In addition, for $T^3 T^3 - T^3 Y$ and $YY - T^3 Y$, i.e., $Z$-exchange between a neutral and a charged particle, a long-distance $r^{-3}$ tail from photon and $Z$-mixing at NLO arises. To the best of our knowledge, this is a completely new long-range potential not discussed previously in the literature.

To make our results accessible, we provide fitting functions that approximate the full numerical Fourier transform of the NLO correction to permille level accuracy. While this paper focuses on the technical computation of the NLO potentials and their behaviour, many phenomenological analyses become possible with the fitting functions at hand. For example, a reanalysis, e.g., of the viable minimal DM candidates is on the cards, both in indirect detection and relic abundance considerations \cite{Cirelli:2007xd}. Similarly, it may also be interesting to assess the impact of the correction on bound state formation, e.g., in the accurate determination of the minimal DM relic abundance \cite{Mitridate:2017izz,Bottaro:2021snn}.

Finally, a comment on the applicability of the results is in order. The results allow the construction of the NLO correction for arbitrary electroweak charged models which obey $\mchi \gg \mZ$. The results are also an integral part of models involving Higgs potentials, such as the MSSM. In this case, however, the correction to the Higgs potentials and depending on the considered gauge, also the tree-level Goldstone and longitudinal gauge boson potentials would need to be worked out. As long as the contribution of Higgs potentials is small, which is typically the case, e.g., in the MSSM \cite{Beneke:2014hja}, our results already provide a good approximation to the full NLO correction also in this case.

\paragraph{Acknowledgements}
I would like to thank M. Beneke and R. Szafron for useful discussions and careful reading of the manuscript. This work was supported in part by the DFG Collaborative Research Centre
``Neutrinos and Dark Matter in Astro- and Particle Physics'' (SFB 1258).

\appendix

\section{Analytic coefficients for asymptotic behaviours}
\label{app:asy}
In Section~\ref{sec:asymain}, we discussed the asymptotic behaviour for $r \to 0 / \infty$ of the NLO correction to the potentials. In this Appendix, the lenghty functions of SM masses and Weinberg angles are provided that appear in this context. The arctan terms below stem from the simplification of real parts of self-energies, e.g., in the gauge-boson mass renormalization.
\subsection{The $r \to \infty$ asymptotics - arbitrary SM representation}

\begin{align}
  F&(\mW,\mZ,\mh)=\frac{151 s_W^2}{360} - \frac{1049 c_W^2}{72} + \frac{80 s_W^2 c_W^2}{3}-8 s_W^4 - \frac{3}{8} \frac{\mh^2}{\mZ^2} + \frac{1}{12} \frac{\mh^4}{\mZ^4} \nonumber \\
  &+\frac{1}{12}\ln \frac{\mW^2}{\mZ^2} - \frac{\ln \frac{\mh^2}{\mZ^2}}{12 \left(\frac{\mh^2}{\mZ^2}-1\right)} \left(- \frac{3}{2} \frac{\mh^2}{\mZ^2} + \frac{\mh^4}{\mZ^4} - \frac{7}{24} \frac{\mh^6}{\mZ^6} + \frac{1}{24} \frac{\mh^8}{\mZ^8}\right) \nonumber \\
  &+\sqrt{\frac{4 \mZ^2}{\mh^2}-1} \arctan\left[\sqrt{\frac{4 \mZ^2}{\mh^2}-1} \right] \left(- \frac{\mh^2}{\mZ^2}+ \frac{1}{3} \frac{\mh^4}{\mZ^4} - \frac{1}{12} \frac{\mh^6}{\mZ^6}\right) \nonumber \\
  &+\sqrt{\frac{4 \mW^2}{\mZ^2}-1} \arctan\left[\frac{\sqrt{\frac{4 \mW^2}{\mZ^2}-1}}{\frac{2\mW^2}{\mZ^2}-1} \right] \left(- \frac{1}{12}- \frac{4}{3} c_W^2 + \frac{17}{3} c_W^4 +4 c_W^6\right)
  \label{eq:asyT3YFfunc}
\end{align}

\begin{align}
  I&(\mW,\mZ,\mh) = \frac{1349 s_W^2}{90} - \frac{s_W^2}{12 c_W^4} - \frac{11 s_W^2}{8c_W^2} + \frac{56 c_W^2 s_W^2}{3} + 8 c_W^4 s_W^2 \nonumber \\
   &+ \frac{3 s_W^2}{8} \frac{\mh^2}{\mW^2} - \frac{(1+c_W^2)s_W^2}{12} \frac{\mh^4}{\mW^4} \nonumber \\
   &+\ln \frac{\mW^2}{\mZ^2} \left(\frac{1}{12} + \frac{17}{4 c_W^2} - \frac{7}{12 c_W^4} - \frac{1}{24 c_W^6} - \frac{3}{4} \frac{\mh^2}{\mW^2} + \frac{1}{4} \frac{\mh^4}{\mW^4} - \frac{1}{24} \frac{\mh^6}{\mW^6}\right) \nonumber \\
   &+ \frac{\ln \frac{\mh^2}{\mZ^2}}{1 - \frac{\mZ^2}{\mh^2}} \left(- \frac{3 s_W^2}{2 c_W^2} + \frac{s_W^2 (1+4c_W^2)}{4 c_W^2} \frac{\mh^2}{\mW^2}+ \frac{7 c_W^6 - 6 c_W^2 -1}{24 c_W^2} \frac{\mh^4}{\mW^4} + \frac{1-c_W^6}{24} \frac{\mh^6}{\mW^6} \right) \nonumber \\
   &+\sqrt{\frac{4 \mW^2}{\mZ^2}-1} \arctan \left[\sqrt{\frac{4 \mW^2}{\mZ^2}-1}\right] \left(\frac{1}{12 c_W^6} + \frac{4}{3 c_W^4} - \frac{17}{3 c_W^2} -4\right) \nonumber \\
   &+\sqrt{\frac{4 \mW^2}{\mZ^2}-1} \arctan \left[\frac{\sqrt{\frac{4 \mW^2}{\mZ^2}-1}}{\frac{2 \mW^2}{\mZ^2}-1}\right] \left(- \frac{1}{12} - \frac{4 c_W^2}{3} + \frac{17 c_W^4 }{3} + 4 c_W^6\right) \nonumber \\
   &+\sqrt{\frac{4 \mW^2}{\mh^2}-1} \arctan \left[\sqrt{\frac{4 \mW^2}{\mh^2}-1}\right] \left(\frac{\mh^2}{\mW^2} - \frac{1}{3} \frac{\mh^4}{\mW^4} + \frac{1}{12} \frac{\mh^6}{\mW^6} \right) \nonumber \\
   &+\sqrt{\frac{4 \mZ^2}{\mh^2}-1} \arctan \left[\sqrt{\frac{4 \mZ^2}{\mh^2}-1}\right] \left(-\frac{\mh^2}{\mZ^2} + \frac{1}{3} \frac{\mh^4}{\mZ^4} - \frac{1}{12} \frac{\mh^6}{\mZ^6} \right)
  \label{eq:asyYYIfunc}
\end{align}

\subsection{The $r \to 0$ asymptotics - arbitrary SM representation}

\begin{align}
  G&(\mW,\mZ,\mt) =- \frac{80 c_W^2}{27}+ \frac{16 c_W^2 (1-4c_W^2)}{9} \frac{\mt^2}{\mW^2} +\frac{c_W^2}{2 s_W^2} \frac{\mt^4}{\mW^4}+ \ln \frac{\mt^2}{\mZ^2}\left(\frac{1}{2 s_W^2} - \frac{13}{9}\right) \nonumber \\
   &+\frac{1}{s_W^2} \sqrt{\frac{4 \mt^2}{\mZ^2}-1} \arctan \left[\frac{\sqrt{\frac{4 \mt^2}{\mZ^2}-1}}{ \frac{2 \mt^2}{\mZ^2}-1}\right] \nonumber \\
   &\hspace{3cm}\times \left(-\frac{17}{18} +\frac{20 c_W^2}{9} - \frac{16 c_W^4}{9} - \left(\frac{7}{18} - \frac{40 c_W^2}{9} + \frac{32 c_W^4}{9}\right) \frac{\mt^2}{\mZ^2} \right) \nonumber \\
   &+\frac{c_W^2}{s_W^2} \ln \frac{\mt^2}{\mt^2 - \mW^2} \left(-1+ \frac{3}{2} \frac{\mt^2}{\mW^2} -\frac{1}{2} \frac{\mt^6}{\mW^6}\right)
  \label{eq:asyYYGfunc}
\end{align}

\begin{align}
  H&(\mW,\mZ,\mh) = \frac{80 c_W^2}{3} - \frac{1}{12 c_W^2} - \frac{1}{2} -8 c_W^2 s_W^2 - \frac{c_W^2}{12} \frac{\mh^4}{\mW^4} \nonumber \\
   &+\frac{c_W^2}{s_W^2} \ln \frac{\mh^2}{\mW^2} \left(\frac{1}{24} \frac{\mh^6}{\mW^6} - \frac{1}{4} \frac{\mh^4}{\mW^4}\right) +\frac{1}{s_W^2} \ln \frac{\mh^2}{\mZ^2} \left(- \frac{1}{24} \frac{\mh^6}{\mZ^6} + \frac{1}{4} \frac{\mh^4}{\mZ^4}\right) \nonumber \\
   &+\frac{1}{s_W^2}\ln \frac{\mW^2}{\mZ^2} \left(\frac{c_W^2}{6} - \frac{1}{24 c_W^4} - \frac{7}{12 c_W^2} + \frac{25}{6} - \frac{3}{4} \frac{\mh^2}{\mZ^2}\right) \nonumber \\
   &+\frac{1}{s_W^2} \sqrt{\frac{4 \mW^2}{\mZ^2}-1} \arctan\left[\sqrt{\frac{4 \mW^2}{\mZ^2}-1}\right] \left(-4c_W^2 + \frac{1}{12 c_W^4} + \frac{4}{3 c_W^2} - \frac{17}{3}\right)\nonumber \\
   &+\frac{1}{s_W^2} \sqrt{\frac{4 \mW^2}{\mZ^2}-1} \arctan\left[\frac{\sqrt{\frac{4 \mW^2}{\mZ^2}-1}}{\frac{2 \mW^2}{\mZ^2}-1}\right] \left(- \frac{1}{12}- \frac{4 c_W^2}{3} + \frac{17 c_W^4}{3} + 4 c_W^6\right) \nonumber \\
   &+\frac{c_W^2}{s_W^2} \sqrt{\frac{4 \mW^2}{\mh^2}-1} \arctan\left[\sqrt{\frac{4 \mW^2}{\mh^2}-1}\right] \left(\frac{1}{12} \frac{\mh^6}{\mW^6} - \frac{1}{3} \frac{\mh^4}{\mW^4} + \frac{\mh^2}{\mW^2}\right) \nonumber \\
   &+\frac{1}{s_W^2} \sqrt{\frac{4 \mZ^2}{\mh^2}-1} \arctan\left[\sqrt{\frac{4 \mZ^2}{\mh^2}-1}\right] \left(-\frac{1}{12} \frac{\mh^6}{\mZ^6} + \frac{1}{3} \frac{\mh^4}{\mZ^4} - \frac{\mh^2}{\mZ^2}\right) 
  \label{eq:asyYYHfunc}
\end{align}



\bibliography{references}


\end{document}